\documentclass[useAMS,usenatbib]{mn2e}
\usepackage{epsfig}
\usepackage{color}
\usepackage{subfigure}

\newcommand{\goodgap}{\hspace{\subfigtopskip} \hspace{\subfigbottomskip}}
\newcommand{\vcx}{{\vec{x}}}

\title[Cosmic shear systematics power spectrum]{The power spectrum of systematics in cosmic shear tomography and the bias on cosmological parameters}

\author[V.F. Cardone et al.]{Vincenzo F. Cardone$^{1}$\thanks{Corresponding author\,: {\tt winnyenodrac@gmail.com}}, Matteo Martinelli$^{2,3,4}$, Erminia Calabrese$^5$, Silvia Galli$^{6,7}$, \and Zhuoyi Huang$^1$, Roberto Maoli$^8$, Alessandro Melchiorri$^8$, Roberto Scaramella$^1$ \\
$^1$I.N.A.F.\,-\,Osservatorio Astronomico di Roma, via Frascati 33, 00040 - Monte Porzio Catone (Roma), Italy \\
$^2$SISSA/ISAS, Via Bonomea 265, 34136, Trieste, Italy\\
$^3$INAF-Osservatorio Astronomico di Trieste, Via G.B. Tiepolo 11, I-34131 Trieste, Italy\\
$^4$INFN, Sezione di Trieste, Via Valerio 2, I-34127 Trieste, Italy\\
$^5$Sub\,-\,department of Astrophysics, University of Oxford, Denys Wilkinson Building, Keble Road, Oxford, OX1 3RH, UK \\
$^6$Institut d' Astrophysique de Paris, UMR 7095\,-\,CNRS, Paris, France \\
$^7$Universit\'e Pierre et Marie Curie, buolevard Arago 98bis, 75014\,-\,Paris, France \\
$^8$Dipartimento di Fisica, Universit\`{a} di Roma "La Sapienza", Piazzale Aldo Moro, 00185 - Roma, Italy \\}

\date{Accepted xxx, Received yyy, in original form zzz}

\begin{document}

\maketitle

\begin{abstract}

Cosmic shear tomography has emerged as one of the most promising tools to both investigate the nature of dark energy and discriminate between General Relativity and modified gravity theories. In order to successfully achieve these goals, systematics in shear measurements have to be taken into account; their impact on the weak lensing power spectrum has to be carefully investigated in order to estimate the bias induced on the inferred cosmological parameters. To this end, we develop here an efficient tool to compute the power spectrum of systematics by propagating, in a realistic way,  shear measurement, source properties and survey setup uncertainties. Starting from analytical results for unweighted moments and general assumptions on the relation between measured and actual shear, we derive analytical expressions for the multiplicative and additive bias, showing how these terms depend not only on the shape measurement errors, but also on the properties of the source galaxies (namely, size, magnitude and spectral energy distribution). We are then able to compute the amplitude of the systematics power spectrum and its scaling with redshift, while we propose a multigaussian expansion to model in a non-parametric way its angular scale dependence. Our method allows to self-consistently propagate the systematics uncertainties to the finally observed shear power spectrum, thus allowing us to quantify the departures from the actual spectrum. We show that even a modest level of systematics can induce non-negligible deviations, thus leading to a significant bias on the recovered cosmological parameters.

\end{abstract}


\section{Introduction}

A plethora of observational evidences (see, e.g., \citealt{W12} and refs. therein and \citealt{Ade13} for recent {\it Planck} results) makes the picture of a spatially flat universe with a subcritical matter content undergoing accelerated expansion a fully accepted paradigm of modern cosmology. What is complementing the cosmic budget and driving the cosmic speed up remains however still largely unknown. Contrary to what typically happens, the problem here is not the absence of a viable solution, but rather the presence of too many candidates playing the game of thrones, with no contenders being so unequivocally superior to the other to be awarded the dark energy crown. Most of the proposed mechanisms, from the classic cosmological constant to dynamical dark energy fluids, from scalar fields with a suitable potential to modified gravity theories (see e.g. \citealt{C11} and refs. therein), are indeed able to nicely fit data probing the background evolution of the cosmos. Although they also make distinct predictions on the growth of perturbations, this latter quantity can still not be traced well enough by the presently available data to allow for a definitive discrimination.

As also witnessed by the conclusions of the Dark Energy Task Force report \citep{DETF}, weak gravitational lensing (also referred to, in the following, as cosmic shear) has emerged as the most promising tool to put some order in the chaotic accumulation of up-to-now viable models. The cosmic shear power spectrum is indeed able to probe both the expansion rate and the growth of perturbations \citep{BS01,HJ08,M08,Huterer:2010hw} thus offering a potentially unique tool to severely constrain cosmological parameters and discriminate among theoretical models. It is therefore not surprising that great efforts have been recently invested in this field on both theoretical studies and observational resources, with many future surveys being planned (e.g., PanSTARRS\footnote{{\tt http://pan-starrs.ifa.hawaii.edu}}, DES\footnote{{\tt http://www.darkenergysurvey.org}}, LSST\footnote{{\tt http://www.lsst.org}} and Euclid\footnote{{\tt http://www.euclid-ec.org}}).

The power of weak lensing as a gravity probe risks however to be strongly diminished by systematic uncertainties, which can cooperate to turn the observed signal into an unfaithful realization of the underlying cosmic one. Such systematics may come from both theoretical issues (such as the linear to nonlinear mapping of the matter power spectrum and the intrinsic alignment of source galaxies) and observationally related problems (as, e.g., the mismatch between the actual redshift distribution and the one inferred from photo\,-\,$z$ and the shear measurement method). A central role in forecasting the precision a future survey can achieve on the cosmological parameters is therefore played by the understanding of the impact of systematics on the cosmological parameters estimation. Moreover, such an analysis would help in identifying which kind of systematics drives the bias on the parameters of interest, thus helping in fixing both the survey strategy and instrumental setup.

Motivated by this consideration, different strategies have been proposed to address this issue. On one hand, much work has been dedicated to estimate the impact of systematics on the shear signal; e.g., \cite{V04} demonstrated how easy it is for systematics to produce spurious B\,-\,modes and make the measured lensing E\,-\,modes departing from the actual ones. This motivated \cite{Man05} to work out a method for testing for the presence of systematics and to apply it to the SDSS galaxy\,-\,galaxy weak lensing data. As first pointed out in \cite{Henk04}, imperfect modelling of the PSF soon appeared to be one of the most likely source of systematics, so that both \cite{St07} and \cite{PH08} examined  to what accuracy the PSF must be modelled  in order  to not bias the shear signal. Since the shear field is reconstructed from the observed shape of galaxies, it is clear that systematics in shape measurement play a significant role as discussed in, e.g., \cite{B10}. All these works are mainly concerned with the impact of systematics on the weak lensing power spectrum, but do not perform a detailed analysis of how they propagate to bias the estimate of cosmological parameters. Not surprisingly, such an issue has then been carefully addressed in the literature under different assumptions about the systematics considered and about their modelling \citep{Hut06,TKH07,KTH08,Kitch09,BH10}. Of particular interest for our aim is the work of Amara \& Refregier (2008, hereafter AR08) who have presented a Fisher matrix analysis to infer the bias on the dark energy equation of state due to the mismatch between theoretical and observed cosmic shear power spectrum caused by the systematics term. A similar but more detailed analysis has also been carried out in \cite{D11}, where a wider set of sources has been taken into account. Both these works, however, assume a parameterized analytical expression for the systematics power spectrum, so that their results hold true only as far as this a priori description is deemed as reliable. A significant step forward has been recently represented by the work of Massey et al. (2013, hereafter M13) where an analytical formalism has been developed to propagate observational uncertainties (coming from errors in shape measurement, PSF correction and CCD defects) on the cosmic shear power spectrum.  A Fisher matrix analysis is then used to set requirements on the amplitude of systematics asking that they do not spoil down the efficiency of the survey in correctly constraining cosmological parameters. These results are then used as an input to the analysis of \cite{Cr13} where a list of requirements on the different systematics is presented.

Most of the works quoted above are mainly concerned with quantifying how systematics alter the cosmic shear power spectrum and hence bias the cosmological parameter determination. Although a careful control of the different sources can help reducing systematics, one should also be ready to deal with the possibility that they could not be completely removed. As such, a model for their power spectrum should be added to the lensing one in the likelihood analysis in order to reduce the bias on the cosmological parameters. Such a model would likely be parametrized by a set of nuisance parameters which inevitably degrade the precision in the estimate of the cosmological ones. Finding a compromise between reducing the bias and improving the precision is a difficult task that is worth to be investigated. The first steps to be addressed to achieve this goal are\,:

\begin{itemize}

\item{find a method to propagate systematics (originating from, e.g., shape measurement and imperfect PSF modelling) on the
finally observed power spectrum;} \\

\item{take into account both the survey setup (PSF, filter transmission curve, magnitude limit) and the properties (size and spectral energy distribution) of the source galaxies;} \\

\item{correctly describe the evolution with redshift of the systematics power spectrum;} \\

\item{model the angular scale dependence of the deviations of the observed shear power spectrum from the lensing one.}

\end{itemize}
The method we present here aims at fulfilling the above four requirements under very general conditions. In order to be as general as possible, we do not make any assumption about the sources of systematics, but only assume that their effect on the shear estimate can be described as a first order deviation as usually assumed in the literature \citep{H06,M07}. Although our method is fully general, we explicitly implement it for the Euclid survey \citep{RB} using realistic assumptions for both the filter transmission curve and PSF wavelength dependence.

The structure of the paper is as follows. In Sect.\,2 we show how systematics can be included in the computation of the cosmic shear power spectrum and describe the multiplicative and additive bias. A general formalism for computing these two terms is presented in Sect.\,3, while the particular case of the planned Euclid survey is considered in Sect.\,4, where we make a step\,-\,by\,-\,step derivation of the systematics power spectrum. Its impact on the determination of the cosmological parameters is investigated in Sect.\,5 through a Markov Chain Monte Carlo likelihood analysis of mock datasets for two different systematics power spectra. A summary of the results and some further considerations are given in the concluding Sect.\,6, while some supplementary material is presented in Appendices A and B.

\section{Observed vs actual shear}

The matter distribution along the line of a sight to a given source alters the shape of the image due to the magnifying effect of the convergence field $\kappa$ and the change in the ellipticity due to the shear $\gamma$. Since neither the intrinsic size nor the source ellipticity are known, one can only resort to statistical methods to get an estimate of both $\kappa$ and $\gamma$. To this end, one first quantifies the shape of an image introducing the second order moments \citep{BS01}

\begin{equation}
Q_{ij}(\lambda) = \frac{\int{{\cal{I}}(\vcx, \lambda) x_i x_j dx_1 dx_2}}{\int{{\cal{I}}(\vcx, \lambda) dx_1 dx_2}} \ \quad \ (i, j = 1, 2) \ .
\label{eq: qij}
\end{equation}
where $(x_1, x_2)$ are Cartesian coordinates with origin in the galaxy centre and ${\cal{I}}(\vcx, \lambda)$ is the 2D energy distribution normalized in such a way that\,:

\begin{equation}
I(\vcx) = \int{{\cal{I}}(\vcx, \lambda) {\cal{T}}(\lambda) d\lambda} \ ,
\label{eq: defint}
\end{equation}

\begin{equation}
F = \int{dx_1 dx_2 \int{{\cal{I}}(\vcx, \lambda) {\cal{T}}(\lambda) d\lambda}}
\label{eq: deflum}
\end{equation}
with $I(\vcx)$ the intensity profile in a waveband with transmission function ${\cal{T}}(\lambda)$ and $F$ the total flux in that filter.

Rather than directly using second order moments, it is quite more common to use the complex ellipticity $\varepsilon(\lambda)$ and the size $R(\lambda)$ defined as  (see, e.g., \citealt{M12})

\begin{equation}
\varepsilon(\lambda) = \varepsilon_1(\lambda) + {\rm i} \varepsilon_2(\lambda) =
\frac{\int{{\cal{I}}(r, \lambda) r^2 {\rm e}^{2 {\rm i} \theta} r dr d\theta}}{\int{{\cal{I}}(r, \lambda) r^2 r dr d\theta}} \ ,
\label{eq: defvarepsM13}
\end{equation}

\begin{equation}
R^2(\lambda) =
\frac{\int{{\cal{I}}(r, \lambda) r^2 r dr d\theta}}{\int{{\cal{I}}(r, \lambda) r dr d\theta}} \ ,
\label{eq: defomegaM13}
\end{equation}
with $(r, \theta)$ cylindrical coordinates in the image plane. These quantities can also be expressed in terms of the moments as

\begin{equation}
\varepsilon(\lambda) =
\frac{Q_{11}(\lambda) - Q_{22}(\lambda) + 2 {\rm i} Q_{12}(\lambda)}{Q_{11}(\lambda) + Q_{22}(\lambda)} \ ,
\label{eq: epsvsqij}
\end{equation}

\begin{equation}
R^2(\lambda) = Q_{11}(\lambda) + Q_{22}(\lambda) \ .
\label{eq: omvsqij}
\end{equation}
The galaxy shape we observe is the outcome of the lensing effect by the intervening matter along the line of sight. Moreover, we do not directly record the image as it is, but after the convolution with the point spread function (PSF) of the observational setup. In the weak lensing limit $(\gamma, \kappa << 1)$, the image shape parameters, after the distortion due to lensing and the PSF convolution, then read

\begin{equation}
R^2(\lambda) = R^{2}_{gal}(\lambda) + R^{2}_{PSF}(\lambda) \ ,
\label{eq: omegashearpsf}
\end{equation}

\begin{equation}
R^{2}(\lambda) \varepsilon(\lambda) = R^{2}_{gal}(\lambda) \varepsilon_{gal}(\lambda)
+ R^{2}_{PSF}(\lambda) \varepsilon_{PSF}(\lambda) \ ,
\label{eq: epsshearpsf}
\end{equation}
where we used the subscript $gal$ ($PSF$) to denote quantities referred to the galaxy after the effect of lensing (the PSF). Since the image we observe is obtained by collecting photons with different wavelengths, we do not actually measure the wavelength dependent ellipticity and size, but rather their values after integration over the filter waveband. However, it is easy to show that the above relations equally hold for the integrated quantities so that, hereafter, we will drop the $\lambda$\,-\,dependence and use $(\varepsilon, R)$ to denote the wavelength integrated ellipticity and size.

Eqs.(\ref{eq: omegashearpsf})\,-\,(\ref{eq: epsshearpsf}) can be combined into a single relation for the reduced shear $g = \gamma/(1 - \kappa)$. To this end, it is convenient to introduce the following auxiliary quantity

\begin{equation}
\xi = R^{2} \varepsilon = Q_{11} - Q_{22} + 2 {\rm i} Q_{12}
\label{eq: defxi}
\end{equation}
and note that (see Appendix A) the transformation rules for $(\xi, \omega)$ under the effect of shear and PSF convolution read

\begin{equation}
\xi = 2 g \omega_0 + \xi_0 + \xi_{PSF}
\label{eq: xiwl}
\end{equation}

\begin{equation}
\omega = \omega_0 + \omega_{PSF}
\label{eq: omwl}
\end{equation}
where quantities with the $0$ subscript refer to the galaxy before the effect of lensing and PSF convolution and hereafter we define $\omega = R^2$. It is then only a matter of algebra to get

\begin{equation}
2 g = \frac{\xi - \xi_{PSF} - \xi_0}{\omega - \omega_{PSF}} \ .
\label{eq: gtrue}
\end{equation}
Eq.(\ref{eq: gtrue}) makes a step further towards the estimate of the reduced shear $g$, but can not still be used because of the term $\xi_0$ which can not be observationally measured. Moreover, since shape measurement methods are not perfect, the observed values $(\xi_{obs}, \omega_{obs})$ could also differ from the actual ones, $(\xi, \omega)$. On the other hand, as we will show later, in order to compute the PSF shape parameters, we need to know not only the PSF intensity profile, but also the spectral energy distribution (SED) of the galaxy. Should this latter not be exactly known, both $(\xi_{PSF}, \omega_{PSF})$ must be replaced by some estimate $(\xi_{est}, \omega_{est})$ which is obtained by integrating the PSF intensity profile over $\lambda$ with a different SED (for instance, the SED of the nearby stars used to infer the PSF profile). Neglecting the intrinsic ellipticity $\xi_0$ (since averages to zero) and explicitly taking into account the difference between the true and estimated PSF shape parameters, we then obtain the following reduced shear estimator\,:

\begin{equation}
2 g_{obs} = \frac{\xi_{obs} - \xi_{est}}{\omega_{obs} - \omega_{est}} \ .
\label{eq: gobs}
\end{equation}
In order to estimate the difference between the actual reduced shear $g$ and the estimated one $g_{obs}$, we will first assume that the following linear relations hold\,:

\begin{equation}
\left \{
\begin{array}{l}
\xi_{obs} = (1 + m_{\xi}) \xi + c_{\xi} \\
 \\
\omega_{obs} = (1 + m_{\omega}) \omega + c_{\omega} \\
\end{array}
\right . \ ,
\label{eq: shapebias}
\end{equation}

\begin{equation}
\left \{
\begin{array}{l}
\xi_{est} = (1 + \mu_{\xi}) \xi + \gamma_{\xi} \\
 \\
\omega_{est} = (1 + \mu_{\omega}) \omega + \gamma_{\omega} \\
\end{array}
\right . \ .
\label{eq: psfbias}
\end{equation}
Although Eqs.(\ref{eq: shapebias}) and (\ref{eq: psfbias}) extend only to first order,  they are actually well motivated. Indeed, all the shape measurement codes (see, e.g., the list in \citealt{Bridle2010}) are designed in such a way to reduce as much as possible the difference between $(\xi_{obs}, \omega_{obs})$ and $(\xi, \omega)$ so that a linear relation is a good approximation. Following the literature, we will refer to $(m_{\xi}, m_{\omega})$ and $(c_{\xi}, c_{\omega})$ as the shape measurement multiplicative and additive bias, respectively. On the other hand, we will show later that the linear parametrization given by Eq.(\ref{eq: psfbias}) also provides an excellent approximation for the relation among the guessed and actual PSF shape parameters for most cases of practical interest.

Inserting Eqs.(\ref{eq: shapebias}) and (\ref{eq: psfbias}) into Eq.(\ref{eq: gobs}), using the weak lensing limit of Eqs.(\ref{eq: xiobs})\,-\,(\ref{eq: omegaobs}) and averaging over a large sample of galaxies, we finally get\,:

\begin{equation}
g_{obs} = (1 + m) g + c
\label{eq: gobsvsgtrue}
\end{equation}
with

\begin{equation}
m = \left \langle \frac{(1 + m_{\xi}) \omega_0}{(1 + m_{\omega}) \omega_0 + (m_{\omega} - \mu_{\omega}) \omega_{PSF} + (c_{\omega} - \gamma_{\omega})}
- 1 \right \rangle
\label{eq: defmulbias}
\end{equation}

\begin{equation}
c = \left \langle \frac{1}{2} \frac{(1 + m_{\xi}) \xi_0 + (m_{\xi} - \mu_{\xi}) \xi_{PSF} + (c_{\xi} - \gamma_{\xi})}{(1 + m_{\omega}) \omega_0 + (m_{\omega} - \mu_{\omega}) \omega_{PSF} + (c_{\omega} - \gamma_{\omega})} \right \rangle \ ,
\label{eq: defaddbias}
\end{equation}
which we will refer to as the multiplicative and additive bias, respectively. It is worth stressing that $(m, c)$ are obtained by averaging over the properties of a sample of galaxies centred on the position in the sky where the shear has to be estimated. Since the values of $(m, c)$ for each individual system depend on the galaxy properties (structural parameters and SED), Eqs.(\ref{eq: defmulbias}) and (\ref{eq: defaddbias}) actually provides only the first order description of the distribution of $(m, c)$ values. One should, however, consider also the width of the distribution since the wide range in galaxy properties can lead to a non negligible spread of the individual values around the mean ones given by Eqs.(\ref{eq: defmulbias})\,-\,(\ref{eq: defaddbias}) above.

\section{The shear power spectrum}

Although Eq.(\ref{eq: gobsvsgtrue}) has been obtained assuming a particular shear estimator, this expression is actually quite general. Indeed, whatever is the estimator used, the observed shear will typically differ from its actual value because of measurement errors and uncontrolled systematics. Should the reconstruction method be efficient enough, one can postulate a simple linear relation between the true and observed values so that we set \citep{H06,M07}

\begin{equation}
\gamma_{obs}(\theta, z) = [1 + {\cal{M}}(z)] \gamma_{lens}(\theta, z) + \gamma_{add}(\theta, z)
\label{eq: gammaobs}
\end{equation}
for the observed shear of a galaxy at redshift $z$ and with position on the sky given by $\theta = (\theta_1, \theta_2)$. Eq.(\ref{eq: gammaobs}) tells us that the observed shear is related to the one due to the lensing effect, denoted $\gamma_{lens}(\theta, z)$, through the multiplicative bias ${\cal{M}}(z)$ and the additive bias $\gamma_{add}(\theta, z)$. Note that we are here assuming that the multiplicative bias does not depend on the position on the sky which is a quite good approximation considering that $m(z) << 1$. Although Eqs.(\ref{eq: defmulbias})\,-\,(\ref{eq: defaddbias}) explicitly refers to the reduced shear $g$, in the weak lensing limit, $\kappa << 1$ so that $g \simeq \gamma$ with great care. We can therefore set ${\cal{M}} = m$ and $\gamma_{add} = c$ and use the above relations to get an estimate of the multiplicative and additive bias.

In order to constrain cosmological parameters, one is actually not interested to the shear value for single galaxies and how it changes with position and redshift, but rather to its power spectrum. We therefore first define the real space correlation functions\,:

\begin{equation}
\xi_{\pm}(\phi, z, z^{\prime}) = \langle \gamma_1(\theta, z) \gamma_1(\theta + \phi, z^{\prime}) \rangle \pm \langle \gamma_2(\theta, z) \gamma_2(\theta + \phi, z^{\prime}) \rangle
\label{eq: defxipm}
\end{equation}
and then compute the power spectrum as\,:

\begin{equation}
{\cal{C}}(\ell, z, z^{\prime}) = \int{\phi d\phi \left [\xi_{+}(\phi, z, z^{\prime}) J_0(\ell \phi) - \xi_{-}(\phi, z, z^{\prime}) J_4(\ell \phi) \right ]}
\label{eq: cldef}
\end{equation}
where $J_n(x)$ is the Bessel function of order $n$. Using Eq.(\ref{eq: gammaobs}) and assuming that the additive systematic term is uncorrelated with the signal, one straightforwardly gets\,:

\begin{eqnarray}
\langle \gamma_i(\theta, z) \gamma_i(\theta + \phi, z^{\prime}) \rangle & = & [1 + m(z) + m(z^{\prime})+ m(z) m(z^{\prime})] \nonumber \\
 & \times & \langle \gamma_{i,lens}(\theta, z) \gamma_{i,lens}(\theta + \phi, z^{\prime}) \rangle \nonumber \\
 & + & \langle \gamma_{i,add}(\theta, z) \gamma_{i,add}(\theta+ \phi, z^{\prime}) \rangle \ ,
\label{eq: gammacorr}
\end{eqnarray}
so that it is only a matter of trivial algebra to substitute this relation into the definition (\ref{eq: defxipm}) and then the result into Eq.(\ref{eq: cldef}) to finally obtain\,:

\begin{equation}
\hat{{\cal{C}}}_{ij}(\ell) = (1 + {\cal{M}}_{ij}) {\cal{C}}_{ij}(\ell)  + {\cal{A}}_{ij}(\ell)
\label{eq: clend}
\end{equation}
where $\hat{{\cal{C}}}_{ij}(\ell)$ is the observed power spectrum between two bins $(i, j)$ centred on $(z_i, z_j)$, respectively, ${\cal{C}}_{ij}(\ell)$ the lensing contribution (evaluated as described in, e.g., \citealt{Hu99}), ${\cal{M}}_{ij} = m(z_i) + m(z_j) + m(z_i) m(z_j)$ the correction due to the multiplicative bias, and ${\cal{A}}_{ij}(\ell)$ the term originating from the additive systematics.
In order to compute the observed power spectrum in Eq.(\ref{eq: clend}), we need a way to estimate both the multiplicative bias ${\cal{M}}_{ij}$ and the additive bias power spectrum $ {\cal{A}}_{ij}(\ell)$ which we will do in the next section.

As a final remark, we warn the reader that, here, we have not taken into account corrections to the theoretical power spectrum due to the intrinsic alignment of the galaxies. While this can be done (see, e.g., \citealt{HS04,Kirk11}), one should introduce further unknown parameters thus weakening the constraints on the cosmological parameters and increasing possible degeneracies among them. Since we are interested in checking the impact of neglecting systematics when forecasting the accuracy on model parameters, we prefer to not include the intrinsic alignment terms in order to consider the most favourable case.

\section{The systematics power spectrum}

Let us now consider how to estimate the $(m, c)$ in a given position on the sky. Eqs.(\ref{eq: defmulbias})\,-\,(\ref{eq: defaddbias}) show that this is only possible if one preliminary knows the set of quantities which we briefly discuss below.

\subsection{Shear measurement bias\,: $(m_{\xi}, c_{\xi}, m_{\omega}, \gamma_{\omega})$}

These quantities are related to the measurement method adopted and the characteristics of the images one is dealing with. Ideally, one should therefore use end\,-\,to\,-\,end simulations to take into account all the features of both the image analysis pipeline and the ellipticity determination software. This is the underlying philosophy inspiring the GREAT challenges \citep{Bridle2010,Kitching2012} which have, however, not yet included all the realistic noise properties so that the results should be taken with some care.

A correct breakdown of the different source terms contributing to the shear measurement bias \citep{Cr13} can, however, help in finding out how the mutiplicative and additive bias should scale with the galaxy size. According to \cite{M12}, three terms mainly contribute to the shear measurement systematics, namely imperfect PSF modelling, imperfect correction for detector effects and shape measurement bias. Using a Taylor expansion based on moments, they get\footnote{Actually, the results in M13 refer to $R_{PSF}/R_{gal}$ averaged over many galaxies, but we have here simplified their formulae to highlight how the shear measurement bias scales with the galaxy size.} ${\cal{M}} = m_2 R_{gal}^{-2} + m_4 R_{gal}^{-4}$ and ${\cal{A}} = a R_{gal}^{-4}$ where $({\cal{M}}, {\cal{A}})$ actually refer to the errors in estimate of $\varepsilon$ and $(m_2, m_4, a)$ depend on the errors on the PSF shape parameters. A different scaling with the galaxy size $R$ has, however, been found by \cite{Mill13} who, based on using the {\it lensfit} code \citep{LensFitI,LensFitII} on simulated images reproducing the characteristics of the CFHTLenS survey \citep{CFHTLenS}, have found a negligible additive bias and a multiplicative bias scaling as $m \propto \exp{(-\alpha R_{lf} \nu)}/\log{\nu}$, with $\nu$ the S/N ratio and $R_{lf}$ a typical galaxy size quantity. A further contribution to the shear measurement bias can also come from the so called {\it noise bias} \citep{MP12} which introduces a deviation of the measured moments from the actual one which correlates with the galaxy ellipticity.

Although some significant steps have been done towards understanding how the shear measurement bias can be modelled, one do expect that the detailed dependence of noise will change according to which shape measurement method one is using with moment\,-\,based methods likely giving different scalings than fitting\,-\,based algorithms. We can nevertheless note that, notwithstanding the details of the shape measurement process, the higher is the S/N ratio of the galaxy and the larger is its size compared to the PSF one, the closer will the inferred shear be to the intrinsic one. We can therefore qualitatively model the shear measurement bias parameters as\,:

\begin{equation}
y = y_{s} \left ( \frac{R_0^2}{R_{PSF}^{2}} \right )^{-\alpha_y} \nu^{-\beta_y}
\label{eq: biasscaling}
\end{equation}
where $y$ stands for one of the quantities $(m_{\xi}, c_{\xi}, m_{\omega}, c_{\omega})$, $y_s$ is scaling quantity and $(\alpha_y, \beta_y)$ are positive numbers fixing the slope of the dependence on the size and $S/N$ ratio. As we will show later, while we have an estimate of the galaxy $R^2$ parameter, we do not know how to set the $S/N$ ratio so that, in this preliminary analysis, we will fix $\beta_y = 0$ for all the $(m_{\xi}, c_{\xi}, m_{\omega}, c_{\omega})$ cases. We are nevertheless left with 12 parameters\footnote{Since $\xi$ is a complex number, $(m_{\xi}, c_{\xi})$ are actually two components quantities thus leading to $2 \times 2 = 4$ quantities to be fixed. Each of them is set assigning the three parameters $(y_s, \alpha_s, f_y)$ thus leading to a total of $2 \times 2 \times 3 = 12$ parameters.}  thus allowing us to explore still a wide range of possibilities. Compared to the above quoted scaling formulae, Eq.(\ref{eq: biasscaling}) typically gives larger biases (unless $\alpha_y$ is unrealistically large) for a given galaxy size, while a further downgrading is introduced because of neglecting the $S/N$ dependence. As a consequence, we expect that our results would overestimate the impact of bias thus making us err on the conservative side.

Eq.(\ref{eq: biasscaling}) is, by definition, only an approximation of how the bias depends on the galaxy properties. As such, it will be affected by a given scatter $\sigma_y = f_y \times y$ (with $f_y$ the fractional uncertainty) thus forcing us to introduce six more parameters. For simplicity, we assume that the actual $y$ value for a galaxy with given $(\omega_0, S/N)$ values may be extracted from a normal distribution centred on $y$ and with variance $\sigma_y^2$. Although simplified, such an approach allows us to realistically estimate the multiplicative and additive bias due to the imperfect recovery of the shear.

\subsection{PSF bias\,: $(\mu_{\xi}, \gamma_{\xi}, \mu_{\omega}, \gamma_{\omega})$}

In order to understand why the estimated PSF $(\xi_{est}, \omega_{est})$ can be different from the actual ones $(\xi_{PSF}, \omega_{PSF})$, it is worth first remembering that these latter are defined as\,:

\begin{equation}
\xi_{PSF} = \int{\xi_{PSF}(\lambda) {\cal{S}}(\lambda) {\cal{T}}(\lambda) d\lambda} \ ,
\label{eq: xipsf}
\end{equation}

\begin{equation}
\omega_{PSF} = \int{\omega_{PSF}(\lambda) {\cal{S}}(\lambda) {\cal{T}}(\lambda) d\lambda} \ ,
\label{eq: omegapsf}
\end{equation}
where the $\lambda$\,-\,dependent shape parameters only depend on the PSF intensity profile  and wavelength dependence, while ${\cal{S}}(\lambda)$ and ${\cal{T}}(\lambda)$ are the galaxy dependent SED and the filter transmission curve. It is worth noting that such a coupling between the PSF and galaxy SED is a consequence of the large width of the Euclid filter. Indeed, should the filter be narrow, one could neglect the $\lambda$ dependence of the PSF and hence avoid the integration. Such an effect was first quantified in \cite{C10}, but here we make a step further explicitly propagating it to the systematics power spectrum considering the actual galaxy SEDs.

The presence of the ${\cal{S}}(\lambda)$ term in Eqs.(\ref{eq: xipsf})\,-\,(\ref{eq: omegapsf}) tells us that errors in the estimate of the PSF shape parameters can also come from an imperfect determination of the galaxy SED. Actually, one can assume that the intensity profile and wavelength dependence of the PSF are very well known so that the main error on the PSF indeed comes from those on the SED. In order to simulate this effect, we will consider two kind of errors. First, we consider the optimistic case that the shape of the SED has been well determined, while a possible redshift mismatch is left with $\Delta z$ randomly extracted from a Gaussian distribution centred on $0.002 (1 + z)$ and variance $0.05 (1 + z)$ and $z$ the actual redshift. As a pessimistic case, we assume that also the shape of the SED has been wrongly estimated. To mimic such errors, we randomly shift by a quantity of order $10\%$ the age of both the bulge and disk so that the colors are approximately the same as expected from a
typical photo\,-\,$z$ code. We denote with $(\xi_{est}, \omega_{est})$ the PSF shape parameters estimated with the wrong SED and fit Eqs.(\ref{eq: psfbias}) for the relation among true and guessed PSF values. It turns out that, within a very good approximation, we can set $\mu_{\xi} = \mu_{\omega} = 0$, while the additive terms $(\gamma_{\xi}, \gamma_{\omega})$ depend on the redshift bin and the SED errors. We have, moreover, checked that the optimistic approach leads to $(\gamma_{\xi}, \gamma_{\omega})$ values which are orders of magnitude smaller than the pessimistic model. We therefore consider only this second option in the following in order to be safely conservative. Finally, to estimate the wavelength dependent PSF shape parameters, we have relied on the simulated PSF model worked out by the Euclid consortium (J. Amiaux, private communication) on the basis of the instrumental setup (telescope and CCD camera) which will be actually used on\,-\,board the satellite. We can therefore be confident that our treatment of the PSF moments is fully realistic.

\subsection{Multiplicative and additive bias}

The two previous paragraphs have demonstrated that the multiplicative and additive bias $(m, c)$ depend on the properties of the galaxy used to infer the shear estimate. In particular, the size and the luminosity (through its impact on the $S/N$ ratio) determine $(m_{\xi}, c_{\xi}, m_{\omega}, c_{\omega})$, while $(\mu_{\xi}, \gamma_{\xi}, \mu_{\omega}, \gamma_{\omega})$ are set by the galaxy SED. We have therefore generated a catalog\footnote{The catalog build up and how each galaxy property have been set are briefly described in Appendix B, while a forthcoming paper will present further details and demonstrating its reliability.} of two components (bulge\,+\,disk) galaxies which can be used as input for the estimate of $(m, c)$ according to the procedure sketched below.

\begin{enumerate}

\item{Pick up a galaxy from the catalog and estimate the intrinsic shape parameters $(\xi_0, \omega_0)$ from the bulge\,+\,disk intensity profile and PSF ones $(\xi_{PSF}, \omega_{PSF})$ according to its composite SED.} \\

\item{Set the shear measurement parameters $(y_s, \alpha_y, f_y)$ and use Eq.(\ref{eq: biasscaling}) and the $\omega_0 = R_0^2$ value to estimate both the mean and the variance of $(m_{\xi}, c_{\xi}, m_{\omega}, c_{\omega})$.} \\

\item{Set $(m_{\xi}, c_{\xi}, m_{\omega}, c_{\omega})_{gal}$ for the given galaxy by randomly sampling from a Gaussian distribution with mean and variance computed above.} \\

\item{Set the PSF bias parameters $(\mu_{\xi}, \gamma_{\xi}, \mu_{\omega}, \gamma_{\omega})$ randomly sampling from a Gaussian distribution with mean and variance estimated as described in Sect.\,4.2.} \\

\item{Repeat steps (i)\,-\,(iv) for ${\cal{N}}$ galaxies and finally estimate $(m, c)$ averaging over the sample values.} \\

\end{enumerate}
The number of ${\cal{N}}$ of galaxies should ideally be very large in order to accurately trace the distribution of the galaxy properties. However, in a realistic application, ${\cal{N}}$ is limited by the number density of the survey and the need to avoid averaging over too large areas in order to not smooth a spatially varying shear signal. For a survey with number density $n_g$, one has to average over an area of radius $\vartheta \simeq ({\cal{N}}/\pi n_g)^{1/2}$ so that the power spectrum can not be estimated for scales larger than $\ell_{max} = \pi/\vartheta$. If we set $n_g = 30 \ {\rm gal/arcmin^2}$ and ${\cal{N}} = 300$, we get $\vartheta \sim 1.8 \ {\rm arcmin}$ leading to $\ell_{max} \simeq 6000$. Actually, since we are interested in tomography, we bin galaxies in redshift according to the binning reported in Table \ref{tab:bins} so that $n_g$ will be smaller than the quoted value in the highest redshift bins. We therefore set ${\cal{N}} = 100$ so that $\ell_{max} \simeq 1900 \ n_g^{1/2}$ which is still reassuringly large also for unrealistically small
$n_g$ values in the highest $z$ bins.

As a final caveat, we warn the reader that we have relied up to now on unweighted moments to derive Eqs.(\ref{eq: defmulbias})\,-\,(\ref{eq: defaddbias}).  Actually, unweighted moments have formally infinite noise (hence $S/N = 0$) requiring to use weighted moments. This introduces a further bias due to the galaxy colour gradients \citep{M12,Sem13} which is not explicitly accounted for in our multiplicative and additive bias estimate. However, \cite{Sem13} have shown that the colour gradient bias can be corrected for leaving a residual effect that can be modelled as a linear perturbation of the observed shape parameters and hence still as in our Eqs.(\ref{eq: shapebias}). We therefore argue that Eqs.(\ref{eq: defmulbias})\,-\,(\ref{eq: defaddbias}) can still be considered approximately valid after calibrating the colour gradient bias.

\begin{table}\footnotesize
\begin{center}
\begin{tabular}{|c|c|c|c|}
\hline
\hline
bin & $z$           & bin  & $z$ \\
\hline
$1$ & $0-0.496$     & $6$  & $1.031-1.163$ \\
$2$ & $0.496-0.654$ & $7$  & $1.163-1.311$ \\
$3$ & $0.654-0.784$ & $8$  & $1.311-1.502$ \\
$4$ & $0.784-0.907$ & $9$  & $1.502-1.782$ \\
$5$ & $0.907-1.031$ & $10$ & $1.782-5.000$ \\
\hline
\hline
\end{tabular}
\caption{Redshift binning used in our analysis. The redshift range of every bin is chosen in such a way that each bin contains 10\% of the galaxies observed by the survey.}
\label{tab:bins}
\end{center}
\end{table}

\subsection{From the real space to the power spectrum}
\label{sec:fromrealtops}
The above procedure allows us to set the values of $(m, c)$ for the shear estimated in a given position $\theta$ on the sky. In order to compute the correlation functions $\xi_{\pm}(\phi, z, z^{\prime})$, we now need to know how the multiplicative and additive bias change as function of the position of the galaxy on the sky. Actually, this is far to be trivial. First, we note that most of the systematics may come from instrument related problems (e.g., charge transfer inefficiency, CCD defects, incorrect flat field correction) or the shape measurement process (which typically depends on the galaxy properties). As such, they are not dependent on $\theta$. On the contrary, one can also think of other more subtle sources of systematics errors which may change on the plane of the sky (e.g., inaccuracies of the photometry calibration depending on the line of sight to the galaxy crossing a region affected by the Milky Way dust or incorrect subtraction of the light coming from a star close in projection to the galaxy of interest). Modelling these effects is hardly possible (if not at all) although some steps have been done in the context of the GREAT10 challenge \citep{Kitching2012}.

In order to parameterize our ignorance, we may phenomenologically proceed setting\footnote{Hereafter, we will only consider the additive bias having assumed that the multiplicative one is scale independent.}

\begin{displaymath}
c_\nu(\theta, z) = \langle c_\nu \rangle(z) F_\nu(\theta, z)
\end{displaymath}
where we have used the notation $c_\nu(z)$ to denote the $\nu$ component of the additive bias $c$ for galaxies in the redshift bin centred on $z$ and $\langle c_{\nu} \rangle(z)$ is estimated as described in Sect.\,4.3. One can then naively obtain the following expression for the real space correlation functions in bins $(i, j)$\,:

\begin{eqnarray}
\xi_{\pm}^{(ij)}(\phi) & = & \langle c_1 \rangle(z_i)  \langle c_1 \rangle(z_j)
\int{F_1(\theta, z_i)  F_1(\theta + \phi, z_j) d\theta} \nonumber \\
 & \pm & \langle c_2 \rangle(z_i)  \langle c_2 \rangle(z_j)
\int{F_2(\theta, z_i)  F_2(\theta + \phi, z_j) d\theta}  \nonumber \ .
\end{eqnarray}
Let us now assume that the systematics errors are isotropic so that they ony depend on $|\theta|$. Using polar coordinates in the integrals, we finally get\,:

\begin{figure*}
\centering
\subfigure{\includegraphics[width=7.5cm]{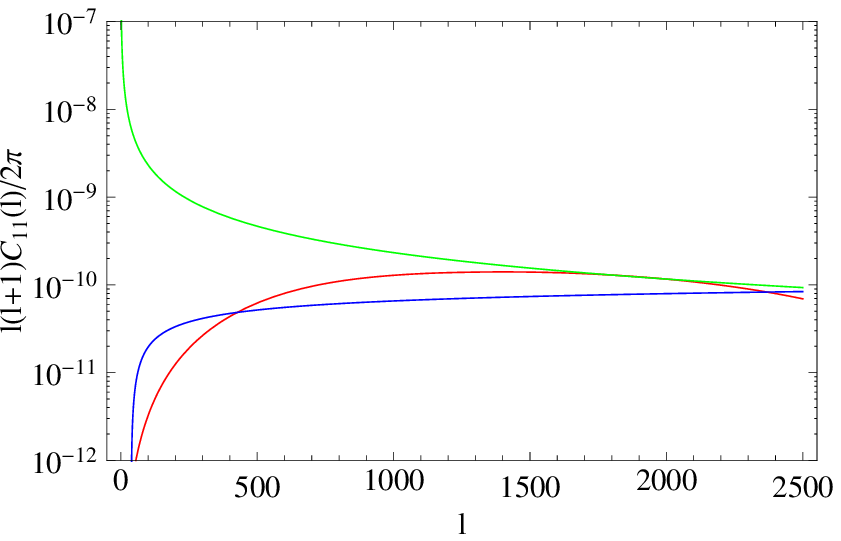}} \goodgap
\subfigure{\includegraphics[width=7.5cm]{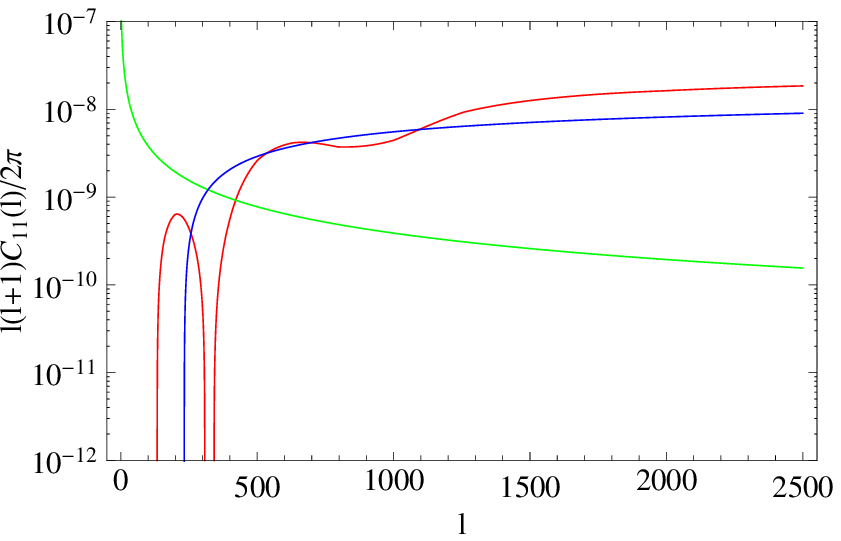}}
\caption{Power spectrum of the additive systematics for the bin combination $(i, j) = (1, 1)$ for the SysA (left) and SysB (right) models with $\vartheta_s = 0.1$ and $0.7 \ {\rm deg}$, respectively. Red, blue and green curves refer to our, Amara \& Refregier (2008) and  Amara et al. (2010) models, respectively. Note that the wiggles in the right panel derive from the choice of using a logarithmic scale on the y-axis, although ${\cal{A}}_{11}(\ell)$ is negative for the plotted model in some $\ell$ ranges.}
\label{fig: psplots}
\end{figure*}

\begin{eqnarray}
\xi_{\pm}^{(ij)}(\eta)/(2 \pi \vartheta_s^2 ) & = & \langle c_1 \rangle(z_i)  \langle c_1 \rangle(z_j) {\cal{F}}_{1}(\eta, z_i, z_j) \nonumber \\
 & \pm & \langle c_2 \rangle(z_i)  \langle c_2 \rangle(z_j) {\cal{F}}_{2}(\eta, z_i, z_j)
\label{eq: xipmadd}
\end{eqnarray}
with $\eta = \phi/\vartheta_s$, $\vartheta_s$ a characteristic scale of the additive systematics and we have introduced the unknown functions

\begin{equation}
{\cal{F}}_\nu(\eta, z_i, z_j) =  \int{F_\nu(\zeta, z_i)  F_\nu(\zeta + \eta, z_j) \zeta d\zeta}
\label{eq: defeffe}
\end{equation}
with $\zeta = | \theta |/\vartheta_s = (\theta_1^2 + \theta_2^2)^{1/2}/\vartheta_s$. The power spectrum ${\cal{A}}_{ij}(\ell)$ can then be computed by inserting Eqs.(\ref{eq: xipmadd})\,-\,(\ref{eq: defeffe}) into Eq.(\ref{eq: cldef}) provided the functions ${\cal{F}}_{\nu}(\eta, z_i, z_j)$ have been given.

To this end, we use a ${\cal{N}}_{F}$\,-\,order multigaussian expansion setting

\begin{equation}
F_\nu(\eta, z_i) =
\sum_{k = 1}^{{\cal{N}}_{F}}{w_{\nu k}(z_i) \exp{ \left [ - \frac{1}{2} \left ( \frac{\eta - \eta_k}{\sigma_k} \right )^2 \right ]}}
\label{eq: fzmge}
\end{equation}
where the weights $w_{\nu k}(z_i)$ are normalized so that

\begin{equation}
\sum_{k = 1}^{{\cal{N}}_{F}}{w_{\nu k}(z_i) \exp{ \left [ - \frac{1}{2} \left ( \frac{\eta_k}{\sigma_k} \right )^2 \right ]}}
= 1 \ ,
\label{eq: normw}
\end{equation}
and we force $(\eta_k, \sigma_k)$ to be smaller than unity so that $\xi_{\pm}^{(ij)}$ will have a starting value set by the amplitudes defined above and then are smoothly truncated for $\phi >> \vartheta_s$. Although Eq.(\ref{eq: fzmge}) has been introduced as useful mathematical tool, it can nevertheless be qualitatively motivated by considering the case of different sources of systematics, each one characterized by an amplitude $\langle c_{\nu} \rangle(z_i)$ and a scale $\vartheta_s$. Without loss of generality, we can scale all the amplitudes with respect to a common value which then plays the role of the amplitude entering Eq.(\ref{eq: xipmadd}). Similarly, the different scales can be expressed as fractions of the largest one thus originating the $\sigma_k$ values in our multigaussian expansion. The only strong assumption is the Gaussian profile for the correlation functions of each component of the full systematics budget. However, the combination of ${\cal{N}}_{F}$ Gaussian functions can mimic quite well a large class of functions if ${\cal{N}}_{F} >> 1$. Eq.(\ref{eq: fzmge}) therefore definitely allows us to explore a wide range of possible behaviours for the systematics correlation functions and hence different power spectrum profiles.

A key role in determining ${\cal{A}}_{ij}(\ell)$ is here played by the scale parameter $\vartheta_s$. Indeed, since the correlation function fades off for $\phi > \vartheta_s$, the power spectrum will be negligibly small for $\ell \le \pi/\vartheta_s$. Although such a feature of the power spectrum is a consequence of the multigaussian expansion, it is also motivated by the physical consideration that it is difficult to think of a source of systematic errors which correlates galaxies far away from each other on the plane of the sky. On the contrary, should the systematics mainly come from something related to the image scale, we should put $\vartheta_s \sim 1 \ {\rm deg}$, this one being the maximum distance two galaxies can have if they are in the same image.

The procedure sketched in Sects.\,4.3 and 4.4 allows us to include in the power spectrum of the additive systematics all the different terms contributing to the final additive bias and and to parametrize our poor knowledge of its scaling with the position through a multigaussian expansion. Fig.\,\ref{fig: psplots} shows ${\cal{A}}_{ij}(\ell)$ for two different models, referred to in the following as SysA and SysB, obtained setting $\vartheta_s = 0.1$ and $0.7 \ {\rm deg}$, respectively, and ${\cal{N}}_F = 10$ components in the multigaussian expansion with randomly generated values of $(\eta_k, \sigma_k, w_{\nu k})$. The amplitude of the systematics depends on the parameters described in Sect.\,4.1 which we set by trial and errors in such a way that the rms percentage deviation of the total power spectrum from the input lensing one is smaller than $0.1 (1) \%$, respectively. For sake of shortness, we only plot the $i = j = 1$ case, i.e., we consider only the autocorrelation for the lowest redshift bin. As a general feature, ${\cal{A}}_{ij}(\ell)$ can be roughly approximated as a power\,-\,law or a logarithmic function for $\ell < \ell_s = \pi/\vartheta_s$, while somewhat flattens and possibly starts oscillating for larger $\ell$.  If we compare to a typical lensing power spectrum, we find that, for reasonable values of the systematics parameters, ${\cal{A}}_{ij}(\ell)$ has a very small impact for $\ell < \ell_s$, while it may be comparable to ${\cal{C}}_{ij}(\ell)$ in the high $\ell$ regime. Such a behaviour could suggest that the systematics have a very negligible impact on the cosmological parameters since they induce a displacement of the observed power spectrum from the actual one only on a quite limited $\ell$ range. However, we warn the reader that the answer to this question depends on the size of the statistical uncertainties and on how the systematics scale with redshift; for this reason, a full analysis is required.

In the same plot, we also show two analytical models for ${\cal{A}}_{ij}(\ell)$ previously used in the literature for testing the impact of systematics. Following AR08, one could set

\begin{equation}
{\cal{A}}_{11}(\ell) = {\cal{A}}_{AR08} \frac{\delta \log{(\ell/\ell_0)} + 1}{\ell (\ell + 1)} \ ,
\label{eq: aar08}
\end{equation}
where we arbitrarily set $\delta = 0.4$. A double power\,-\,law function has been proposed in \cite{ARP10} setting\,:

\begin{equation}
{\cal{A}}_{11}(\ell) = {\cal{A}}_{A10} \left ( 1 + \frac{\ell}{\ell_0} \right )^{\beta_2 - \beta_1} \left ( \frac{\ell}{\ell_0} \right )^{\beta_1} \ ,
\label{eq: am12}
\end{equation}
with $(\beta_1, \beta_2) = (-1.5, -3.0)$ as fiducial values. In order to make a meaningful comparison with our power spectrum, we set the two parameters $(\ell_0, {\cal{A}}_k)$ (with $k = AR08, A10$) so that all the models match at $\ell_s = \pi/\vartheta_s$ and have the same value for the variance of the power spectrum, defined as \citep{AR08}

\begin{equation}
\sigma_{sys}^2 = \frac{1}{2 \pi} \int_{\ell_{min}}^{\ell_{max}}{|{\cal{A}}_{11}(\ell)| \ell (\ell + 1) d\ln{\ell}} \ ,
\label{eq: defsigmasys}
\end{equation}
with $(\ell_{min}, \ell_{max}) = (10, 10000)$. Note that this quantity is first estimated from our model so that its value depends on the shear measurement and PSF bias parameters.

It is clear that our model is comparable to the AR08 and \cite{ARP10} ones in the $\ell > \ell_s$ regime, while it predicts a definitely smaller power spectrum for smaller $\ell$ values. This is a consequence of the multigaussian expansion which introduces a smooth truncation of the systematics on scales larger than $\vartheta_s$. Since such a cutoff is physically motivated, we are confident that the mismatch with the AR08 and \cite{ARP10} models is not a worrisome problem, but rather a nice improvement.

As a final remark, we stress that the comparison shown in Fig.\,\ref{fig: psplots} should not be overrated. Indeed, our aim is not to reproduce any given analytical formula. On the contrary, we model the additive systematics power spectrum ${\cal{A}}_{ij}(\ell)$ in a non parametric way as the product of a redshift dependent amplitude $\langle c(z_i) \rangle \langle c(z_j) \rangle$, computed following the steps detailed in Sect.\,4.3, and a non parametric scale dependent profile derived starting from the multigaussian expansion described before. From this point of view, our approach is similar to the one employed in \cite{M12} where the scale dependence of the systematics power spectrum is again modelled non parametrically although relying on a different method based on the form filling approach \citep{Kitch09}.

\section{Bias on cosmological parameters}

The  procedure described in the previous sections allows us to estimate both the redshift dependent multiplicative bias ${\cal{M}}(z)$ and the additive power spectra ${\cal{A}}_{ij}(\ell)$, so that the observed power spectra $\hat{{\cal{C}}}_{ij}(\ell)$ can be naively computed provided that a reference cosmological model is set as an input for the lensing power spectrum. We can therefore evaluate the impact of the systematics on cosmological parameters by using a two steps approach. First, we generate a mock dataset including the systematics terms, i.e. we use as an input the power spectra computed according to Eq.(\ref{eq: clend}). We then perform a Markov Chain Monte Carlo (MCMC) analysis, fitting the mock data with the theoretical unbiased shear power spectrum, i.e. setting ${\cal{M}}(z) =  {\cal{A}}_{ij}(\ell) = 0$. The comparison between the input parameters and the recovered ones will highlight to which extent the cosmological parameters are biased by the presence of uncorrected systematics. It is worth stressing that we are considering here the worst case scenario where one is fully unaware of the presence of systematics. A step forward would be represented by a likelihood analysis fitting both for the cosmological parameters and a nuisance set of quantities introduced to model the systematics. Up to now, such a modelling is unavailable so that we will investigate this second step in a forthcoming paper.

Our fiducial cosmological model is the Lambda Cold Dark Matter ($\Lambda$CDM) standard scenario where the dark energy component is described by the cosmological constant $\Lambda$ with a constant equation of state $w=-1$.
 We first study the impact of systematics on the $\Lambda$CDM parameters, and then assume a CPL parametrization for dark energy \citep{CP01,L03}. In this model the dark energy equation of state reads $w(a) = w_0 + w_a (1 - a)$, where $w_0$ and $w_a$ are the present day value of the equation of state and its derivative and $a$ is the scale factor.

Before discussing the results of such a test, it is worth comparing it to the previous works in the literature. As yet said in the introduction, the impact of systematics on the cosmological parameters was first addressed in AR08 through a Fisher matrix analysis. Although important as a first step, the AR08 results are affected by two shortcomings. First, their systematics power spectrum was given a priori without any connection with what is actually originating the systematics. Although Fig.\,\ref{fig: psplots} shows that the AR08 model gives a reasonable approximation of our ${\cal{A}}_{ij}(\ell)$ profile, its parameters were set by hand so that the inferred results can hardly be related to the actual amplitude of the multiplicative and additive bias. On the contrary, we determine both ${\cal{M}}(z)$ and ${\cal{A}}_{ij}(\ell)$ by taking into account the characteristics of both the Euclid survey and the source galaxies, while the shape measurement uncertainties and PSF reconstruction errors are correctly propagated.

A similar underlying philosophy drives the work of M13 where the errors from shape measurement and charge transfer inefficiency are propagated on the final estimate of the shear. Our approach shares with M13 the adoption of first order relations between input and measured moments and a non parametric description of the scale dependence of the additive systematics power spectrum. However, we use a different way to propagate shear and PSF uncertainties on the amplitude of the systematics, in order to take into account the details of the galaxy shape parameters and SED. As a further difference, we are not interested in budgeting the impact of each kind of systematics (see also \citealt{Cr13}), but rather in an overall analysis. These differences are mainly motivated by the targets of the two works. While M13 aims at defining requirements to reduce the different sources of systematics, we are here more concerned with estimating the systematics power spectrum taking into account the characteristics of the target galaxies and the survey setup. From this point of view, our work can be considered as a sort of cross check of their results. Should we find a bias on the cosmological parameters when the systematics amplitude is larger than their limits notwithstanding the different modelling of the systematics power spectrum, their requirements on the systematics would be enforced.

An important difference with both AR08 and M13 is related to how we infer the bias on the cosmological parameters. In the cited papers, the authors add the systematics power spectrum to the input one and relies on a Fisher matrix analysis to estimate the final bias. On the contrary, we here rely on mock data and the standard MCMC technique to constrain cosmological parameters. On  one hand, we are forced to make this choice because we cannnot work out an analytical likelihood term for the systematics given the way we have computed the related terms. On the other hand, although user friendly, the Fisher matrix analysis can lead to biased constraints if the likelihood function is not well approximated by a multi Gaussian distribution close to its maximum. Moreover, the numerical computations typically needed to compute the Fisher matrix elements can miss some pathological behaviour of the likelihood derivative with respect to the model parameters thus leading to unreliable results (see, e.g., \citealt{Mat10} for a  discussion and a comparison of the two approaches).

\subsection{Mock datasets}

We simulate cosmic shear power spectra with and without systematics contributions, and we consider a flat $\Lambda$CDM cosmological model as the fiducial input cosmology. We set the cosmological parameters to the WMAP7 mean marginalized values \citep{WMAP7} : $\Omega_{b} h^2 = 0.02258$ and $\Omega_{c} h^2 = 0.1109$ for the baryon and CDM physical matter densities, $n_s = 0.963$ for the scalar spectral index, $A_s = 2.43 \times 10^{-9}$ for the scalar amplitude as evaluated at a pivot scale $k = 0.002 \ {\rm Mpc^{-1}}$. Furthermore, we use the value of the  Hubble constant, $H_0 = 70.767 \ {\rm km/s/Mpc}$, derived from the marginalized mean value of the angular size of the sound horizon at last scattering surface, $\theta=0.010388$.

The weak lensing dataset has been simulated according to the specifications in Table\,\ref{tabeuclid}, in agreement with what is expected for the Euclid survey \citep{RB}. The mission will observe $n_g \simeq 30 \ {\rm gal/arcmin^2}$ over an area $\Omega = 15000 \ {\rm deg^2}$ and the same redshift distribution adopted to generate the galaxy catalog used to quantify the systematics. The large galaxy number density and the wide area observed will allow Euclid to perform a tomographic reconstruction of the weak lensing signal. We therefore divide the redshift space in ten bins with the same ranges assumed in the systematics power spectra computation (see Table \ref{tab:bins}). We then generate the shear power spectrum ${\cal{C}}_{lens}(\ell, z, z^{\prime})$ following the prescription described in \cite{fdebe011}. The $1 \sigma$ uncertainties on the convergence power spectrum can be computed as in \cite{Cooray:1999rv}\,:

\begin{equation}
\label{sigmaconv}
\sigma_{\ell}=\sqrt{\frac{2}{(2\ell+1)f_{sky}\Delta_{\ell}}}\left[P(\ell)+\frac{\gamma_{rms}^2}{n_{gal}}\right]~,
\end{equation}
where $\Delta_{\ell}$ is the width of the $\ell$\,-\,bin used to generate the data. Here, we simply choose $\Delta_\ell= 1$ for the full range considered, i.e. $\ell \in [2,2500]$. Note that we do not consider higher $\ell$ in order to  avoid both the modelling of nonlinear effects and the impact of baryons on the lensing power spectrum \citep{Sem11}, which can cause a systematic offset between the theoretical power spectrum and the actual one.

\begin{table}
\begin{center}
\begin{tabular}{ll}
\hline
\hline
n$_{\rm gal}$ (arcmin$^{-2}$) & $30$ \\
redshift & $0<z<5$\\
Sky Coverage(deg$^2$)  & $15000$\\
$\gamma_{\rm rms}$ & $0.30$\\
\hline \hline
\end{tabular}
\caption{Specifications for the Euclid like survey considered in this paper. The table shows the number of galaxies per square
arcminute ($n_{gal}$), redshift range, sky coverage and intrinsic ellipticity ($\gamma^2_{rms}$) per component.} \label{tabeuclid}
\end{center}
\end{table}

We compute a mock dataset for Euclid introducing the effect of systematics as in Eq.(\ref{eq: clend}) focusing on the two SysA and SysB models shown in Fig.\,\ref{fig: psplots} and described in Sect. \ref{sec:fromrealtops}. Note that the Euclid's field of view will roughly be $0.7 \times 0.7 \ {\rm deg}^2$ so that the choice $\vartheta_s = 0.7 \ {\rm deg}$ made for the SysB model means that we are assuming that the systematics are related to some undetected phenomenon acting on the scale of the image. Similarly, $\vartheta_s = 0.1 \ {\rm deg} = 6 \ {\rm arcmin}$ is typically the smallest scale where the shear correlation function is reliably measured so that this choice for model SysA mimics the optimistic assumption that no significant systematics are present on larger scales. We also remind the reader that the rms percentage deviation of the systematics is of order $0.1\%$  $(1\%)$ for model SysA (SysB).
By considering these two cases, we are therefore investigating the impact of systematics under both optimistic and pessimistic assumptions on their impact.

\subsection{Results}

As anticipated, we discuss here whether the presence of uncorrected systematics in the shear power spectrum introduces a significant bias in the inferred cosmological parameters. To this end, we perform an MCMC analysis based on a modified version of the publicly available package \texttt{cosmomc} \citep{Lewis:2002ah} and a modified version of the weak lensing module developed by \cite{cosmos}. We check for convergence of the chains through the \cite{GR92} test. We explore a $\Lambda$CDM
model sampling over the following parameters\,:

\begin{table}
\begin{center}
\begin{tabular}{|l|c|c|c|}
\hline

 & no sys & SysA & SysB  \\
\hline
$\Lambda$CDM   & $0.006$ & $94$ & $30276$\\
$w_0w_a$CDM & $0.23$ & $57$ & $29911$\\
\hline
\hline
\end{tabular}
\caption{Best fit $\chi^2$ values for the performed analysis with $\Lambda$CDM and $w_0w_a$CDM.}
\label{tab:like}
\end{center}
\end{table}

\begin{displaymath}
{\bf p} = (\Omega_c h^2, H_0, n_s, A_s)
\end{displaymath}
adopting flat priors on all of them. We also consider an extended $w_0 w_a$CDM model that includes, besides the above mentioned four $\Lambda$CDM parameters, two additional parameters, $w_0$ and $w_a$ (still assuming flat priors on these latter). We do not vary the baryon density $\Omega_b h^2$, but rather keep it fixed to the WMAP7 value. We have indeed verified that degeneracies between $\Omega_b h^2$ and other parameters are so large that it is nearly impossible to constrain this parameter if it is let free to vary. We have verified that imposing a strong gaussian prior, at the 1\% level (compatible with the measurements performed by CMB experiments such as Planck), on this parameter, yields to results that are qualitatively similar to the ones obtained by fixing its value. We thus present the results derived in this second case.

 As already mentioned above, the analysis performed here represents the first step in a detailed investigation of the impact of systematics in weak lensing surveys. In particular, we here want to verify the validity of a zeroth order approach, where one erroneously assumes that systematics have been reduced to such a small level that the residual ones can be neglected in the fitting procedure. In other words, we  here investigate to which extent such (somewhat optimistic) assumption would bias the constraints on cosmological parameters in presence of unaccounted systematics. A future further step forward will be represented by the inclusion of a parametrized model for the systematics, i.e. a parameterized profile for ${\cal{M}}(z)$ and ${\cal{A}}_{ij}(\ell)$, in the MCMC procedure. Should such a model be detailed enough, one could be able to avoid any bias on the cosmological parameters. As a downside however, the more the model is detailed, the larger it is the set of nuisance parameters describing it. This might open new degeneracies  that could weaken the constraints on cosmological parameters. Investigating this point will be addressed in a forthcoming work.

\begin{figure}
\begin{center}
\hspace*{-1cm}
\begin{tabular}{cc}
\includegraphics[width=7cm]{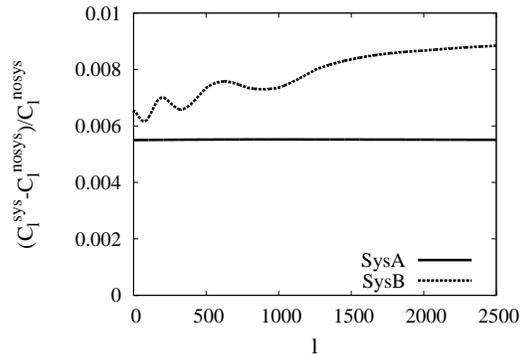}
 \end{tabular}
\caption{Relative difference between the fiducial shear power spectrum and the same spectrum when systematics are included.}
\label{fig:sysplot}
\end{center}
\end{figure}

\subsubsection{Effects on fit statistics}

The MCMC procedure allows to search for the model that best fits the mock dataset, but does not ensure that the match between this model and the data is actually good in general terms. This is clearly shown comparing how the maximum likelihood value (or, equivalently, the $\chi^2$ for the best fit model) worsens when including systematics in the mock data. As can be read from Table\,\ref{tab:like} for both the $\Lambda$CDM and $w_0w_a$CDM models, the best fit $\chi^2$ is roughly two (four) orders of magnitude larger for the SysA (SysB) models compared to the ideal case with no systematics, even though the three fits have the same number of degrees of freedom.

The order of magnitude increase of the best fit $\chi^2$ can be qualitatively explained looking at Fig.\,\ref{fig:sysplot}, where the relative difference between the fiducial power spectrum and the systematics plagued ones is shown for both the SysA and SysB cases. The negligibly small additive power spectrum ${\cal{A}}_{ij}(\ell)$ of the SysA model makes the observed lensing spectrum $\hat{{\cal{C}}}_{ij}(\ell)$ differ from the fiducial one mainly because of a multiplicative term. Should this latter be constant with $z$, the fit could be adjusted by scaling one the of the parameters such as, e.g., the matter density $\Omega_m$. However, the SysA multiplicative bias mildy depends on  redshift, so that such scaling is unable to fully compensate the effect of the systematics, thus leading to a small increase of the best fit $\chi^2$. Moreover, although hardly visible in Fig.\,\ref{fig:sysplot}, the contribution of the additive power spectrum introduces some very small oscillations in the observed power spectrum which can not be reproduced by the best fit model. The extremely high sensitivity of Euclid makes this oscillation non-negligible, thus motivating the rise of the $\chi^2$ value with respect to the case with no systematics. Things get even worse for the SysB case, where the additive bias ${\cal{A}}_{ij}(\ell)$ can be comparable to the fiducial lensing power spectrum in the high $\ell$ regime. As a consequence, both the amplitude and the profile of the observed power spectrum differs from the fiducial one in a way that can no more be compensated by the shift in the cosmological parameters. The very small error bars predicted for such a good quality survey as the Euclid one then boosts the $\chi^2$ to spectacularly large values even if the rms deviation introduced by the systematics is still as small as $\sim1\%$ for the already pessimistic SysB case.

\begin{figure}
\begin{center}
\hspace*{-1cm}
\begin{tabular}{cc}
\includegraphics[width=0.5\textwidth]{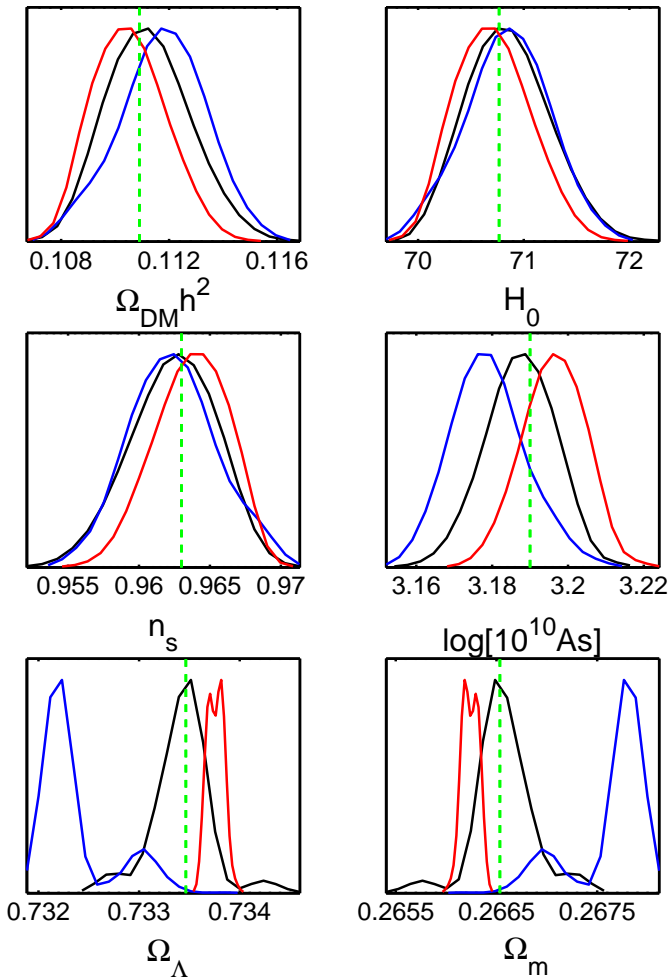}
 \end{tabular}
\caption{Marginalized 1D likelihood profiles for a $\Lambda$CDM model. The upper four plots refer to sampled parameters, while the bottom two refer to derived ones. We show results for Euclid mock data assuming no Sys (black), SysA (red) and SysB (blue) cases. The green  dashed lines show the fiducial cosmological parameters.}
\label{fig: 1Dplot}
\end{center}
\end{figure}

\begin{figure}
\begin{center}
\hspace*{-1cm}
\begin{tabular}{cc}
\includegraphics[width=0.5\textwidth]{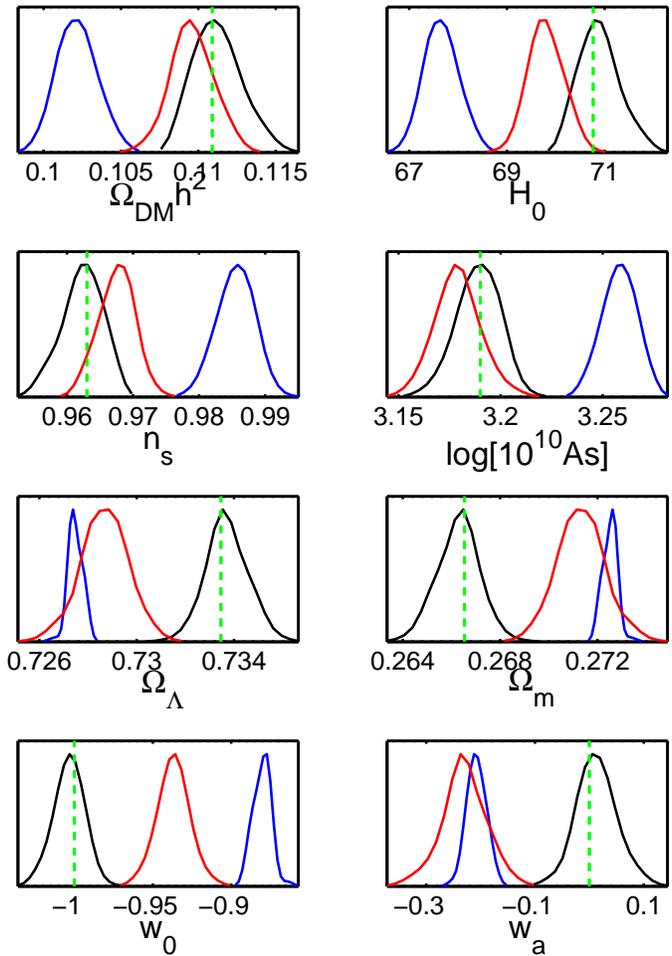}
 \end{tabular}
\caption{Same as Fig.\,\ref{fig: 1Dplot} but for the $w_0w_a$CDM model, where the two parameters $(w_0, w_a)$ are added to the fit.}

\label{fig: cplplot}
\end{center}
\end{figure}

Should such large $\chi^2$ values be found also when analyzing real data,  they could be read as a hint that systematics are present in the data, thus highlighting the need for an analysis that explicitly takes them into account through a suitable parametrized description. However, we have currently only considered a limited set of cosmological models and systematics power spectra. It is therefore premature to draw any general lesson from the above $\chi^2$ values. Moreover, high values of the $\chi^2$ in real data could rather be the hint of the need of a different or more complicated theoretical model to describe the data rather than the symptom of ignored systematics.
 As a conservative conclusion, we can nevertheless notice that the smallness of the statistical errors allows to detect the presence of systematics even if their impact on the observed power spectra is as low as $0.1 \%$ of the underlying cosmological signal.


As a further instructive remark, we would like to stress that the high $\chi^2$ values could only be unveiled thanks to our full MCMC analysis of the mock dataset. Should we have relied on the usual Fisher matrix approach, we would have obtained a legitimate estimate of the bias on the cosmological parameters based on the assumption that the likelihood can be approximated as a Gaussian around the fiducial model. However, we could not have verified if the latter provides a good match to the data. We therefore warn the reader to not overrate the conclusions based on Fisher matrix estimate of the bias since they can be flawed by this mismatch between model and actual data.

\subsubsection{Effect on parameters estimation}

We are now interested in quantifying the bias on the cosmological parameters due to neglecting the systematics in the likelihood analysis.

\begin{table*}
\begin{center}
\begin{tabular}{|l|c|c|c|c|c|c|}
\hline
Parameter & Fiducial & no sys & SysA & $|\Delta/\sigma|$ & SysB & $|\Delta/\sigma|$ \\
\hline
\hline
\hline
$\Omega_m$ & $0.26654$& $0.26658\pm 0.00028$     &$0.266240\pm 0.000080$ &$-3.6$ &$0.26765\pm 0.00034    $&$3.2 $\\
$n_s$&0.963&$0.9624\pm 0.0029$                   &$0.9637\pm 0.0025$     &$0.28$ &$0.9623\pm 0.0029  $&$-0.23 $\\
$log[10^{10}A_s]$&$3.19$&$3.1873\pm 0.0089$      &$3.1962\pm 0.0080$     &$0.77$ &$3.1789\pm 0.0094 $&$-1.2 $\\
$H_0$&$70.77$&$70.85\pm 0.37$                   &$70.71\pm 0.34$        &$-0.17$&$70.86\pm 0.38 $   &$0.24 $\\
$\Omega_\Lambda$ &  $0.73347$& $0.73342\pm 0.00028$     &$0.733760\pm 0.000080$ &$3.6$ &$0.73235\pm 0.00034    $&$-3.2 $\\
\hline
\end{tabular}
\caption{Mean marginalized values and standard deviations of the most relevant cosmological parameters for the $\Lambda$CDM model in the no Sys (second column), SysA (thrid column) and SysB (fifth column) cases. The first column shows the fiducial model, while the fourth (sixth) column shows the bias, defined as the difference ($\Delta$) between the mean marginalized value of the SysA (SysB) case and the fiducial value,  divided by the $1 \sigma$ uncertainty of the SysA (SysB) case.}
\label{tab:shiftLCDM}
\end{center}
\end{table*}

\begin{table*}
\begin{center}
\begin{tabular}{|l|c|c|c|c|c|c|}
\hline

Parameter & Fiducial &no sys & SysA & $|\Delta/\sigma|$ & SysB & $|\Delta/\sigma|$ \\
\hline
\hline

$\Omega_m$ & $0.26654$& $0.26629\pm 0.00084$     &$0.27132\pm 0.00095$   &$5.0$    &$0.27252\pm 0.00030    $&$20.5 $\\
$n_s$&0.963&$0.9623\pm 0.0031$                   &$0.9675\pm 0.0027$     &$1.7$    &$0.9856\pm 0.0028  $&$8.1$\\
$log[10^{10}A_s]$&$3.19$&$3.189\pm 0.010$      &$3.179\pm 0.011$        &$-0.96$  &$3.2582\pm 0.0084 $&$8.1 $\\
$H_0$&$70.77$&$70.90\pm 0.41$                   &$69.80\pm 0.36$        &$-2.7$   &$67.66\pm 0.35 $   &$8.9 $\\
$\Omega_\Lambda$ & $0.73347$& $0.73371\pm 0.00084$     &$0.72868\pm 0.00095$   &$5.0$    &$0.72748\pm 0.00030    $&$-20.5 $\\
$w_0$            & $-1.0$    & $-1.0034\pm0.0096$ & $-0.937\pm0.010$     & $6.2$   & $-0.8803\pm0.0054$ & $22$\\
$w_a$            & $0.0$     & $0.013\pm0.036$ & $-0.230\pm0.040$        & $5.8$   & $-0.208\pm0.018$ & $-12$\\
\hline
\end{tabular}
\caption{Same as Table\,\ref{tab:shiftLCDM} but for the $w_0w_a$CDM model.}
\label{tab:shiftwzwa}
\end{center}
\end{table*}

As a first case, we consider the $\Lambda$CDM model. We first check that we recover the fiducial cosmological parameters when fitting a mock dataset without systematics (\emph{no sys} in the following), e.g. obtained setting  both the multiplicative bias and the power spectrum of the additive systematics to zero. We then run two further cases adding the SysA and SysB systematics to the lensing power spectra before generating the mock datasets.

Fig.\,\ref{fig: 1Dplot} shows the constraints on the most relevant cosmological parameters. As it can be noticed, we correctly recover the fiducial values of the cosmological parameters when no systematics are present, while results are clearly biased in the other cases considered. Furthermore, likelihood profiles appear to be slightly non-gaussian in some cases, specially for the derived parameters ($\Omega_m, \Omega_{\Lambda}$, matter and dark energy density respectively). As a way to quantify the observed bias, we report in Table\,\ref{tab:shiftLCDM} the mean marginalized values and the standard deviations for the most relevant parameters and the corresponding relative bias $\Delta/\sigma$ where $(\Delta, \sigma)$ are the shift from the fiducial model and the standard deviation, respectively. We underline here that due to the slight non-gaussianity of the likelihood distributions, the standard deviations do not necessarily correspond to the 68\% c.l. bounds. Different definitions of the uncertainty $\sigma$ might thus lead to slightly different quantifications of the bias. However, we verified these changes do not affect our overall conclusions.

Following M13, we can deem as {\it biased} a parameter if $\Delta/\sigma$ exceeds the value $\Delta/\sigma > 0.31$ because of the presence of systematics. As Table\,\ref{tab:shiftLCDM} shows, the derived parameters $(\Omega_m, \Omega_{\Lambda})$ are severely biased for both the SysA and SysB cases, while the amplitude $A_s$ of the perturbations is shifted by a small but yet non-negligible amount. On the contrary, both the slope of the primordial power spectrum $n_s$ and the Hubble constant $H_0$ do not significantly shift from the fiducial value, so that their determination is robust against the impact of systematics.  Although we used a different approach to estimate $\Delta/\sigma$, it is nevertheless worth noting that this conclusion is consistent with that of M13. They indeed find that, in order for the systematics to not bias the estimate of cosmological parameters, the two following conditions must be fulfilled\,:

\begin{displaymath}
\bar{{\cal{M}}} < 4 \times 10^{-3} \ \ , \ \ \bar{{\cal{A}}} \le 1.3 \times 10^{-12} \ \ ,
\end{displaymath}
having defined

\begin{displaymath}
\bar{{\cal{M}}} = \frac{\sum_{ij}{\int_{\ell_{min}}^{\ell_{max}}{\left | {\cal{M}}_{ij}(\ell) \right |^2 \ell^2 d\ln{\ell}}}}
{\sum_{ij}{\int_{\ell_{min}}^{\ell_{max}}{ \ell^2 d\ln{\ell}}}} \ ,
\end{displaymath}

\begin{displaymath}
\bar{{\cal{A}}} = \frac{\sum_{ij}{\int_{\ell_{min}}^{\ell_{max}}{\left | {\cal{A}}_{ij}(\ell) \right |^2 \ell^2 d\ln{\ell}}}}
{\sum_{ij}{\int_{\ell_{min}}^{\ell_{max}}{ \ell^2 d\ln{\ell}}}} \ ,
\end{displaymath}
where the sum runs over the different bins combinations\footnote{Note that, since we have assumed the multiplicative bias is scale independent, $\bar{{\cal{M}}}$ reduces in our case to a simple average over the redshift bins.} and $(\ell_{min}, \ell_{max}) = (10, 10000)$. For the SysA model, we get

\begin{displaymath}
\bar{\cal{M}} = 4.2 \times 10^{-3} \ \ , \ \ \bar{{\cal{A}}} = 5.1 \times 10^{-16} \ \ ,
\end{displaymath}
while it is

\begin{displaymath}
\bar{\cal{M}} = 4.5 \times 10^{-3} \ \ , \ \ \bar{{\cal{A}}} = 3.2 \times 10^{-14} \ \ ,
\end{displaymath}
for the SysB systematics. In both cases, the multiplicative bias is larger than the M13 limit, confiming that the bias on  cosmological parameters is expected also from this kind of criterion. Furthermore, such a conclusion is also qualitatively consistent with AR08 suggesting that it is the multiplicative systematics which plays the most important role in biasing the estimate of cosmological parameters. One could naively be surprised that $\Delta/\sigma$ takes such large values considering that, for both systematics models, the condition on $\bar{{\cal{A}}}$ is fulfilled, while the one on $\bar{{\cal{M}}}$ is only mildly violated. However, one should be aware that the M13 limits have been obtained based on a Fisher matrix approach. Since this method is known to underestimate the limits on the bias, it is not surprising that our MCMC fitting of mock (but yet realistic) data gives larger bias values. We therefore recommend to rely on such a method to rederive conservative limits on the $(\bar{{\cal{M}}}, \bar{{\cal{A}}})$ parameters, which is outside our aims here.

As a second test, we now analyze the bias on cosmological parameters for a $w_0w_a$CDM model, allowing the dark energy equation of state to deviate from the constant cosmological constant value $w = -1$ using the CPL parametrization.  Fig.\,\ref{fig: cplplot} shows the constraints on the most relevant cosmological parameters, while Table\,\ref{tab:shiftwzwa} shows the shifts between the recovered parameters and the input fiducial model. While this is consistent with the above expectations based on the $\bar{\cal{M}}$ values, it is worth noticing how the shifts are now significantly larger than in the $\Lambda$CDM case. This is due to the fact that increasing the number of free parameters allows a better fit of the mock datasets that include systematics. Leaving $w_0$ and $w_a$ free to vary does in fact improve the best fit $\chi^2$ shown in Table\,\ref{tab:like}. However, this has the downside that, due to the correlations between parameters, best fitting models can  be determined by extremely more biased cosmological parameters.

\section{Conclusions}

Cosmic shear tomography has emerged as one of the most promising ways to help cosmologists to end the debate on whether GR based quintessence models or modified gravity scenarios are the best candidate to solve the dark energy rebus. The possibility to probe both the background evolution and the growth of structures makes the shear power spectrum an ideal tool to break the degeneracies between the two competing proposals. It is therefore mandatory to be sure that nothing intervenes to degrade this possibility. The next\,-\,to\,-\,come weak lensing surveys are designed to reduce to a negligible level the statistical errors so that the main remaining task is to take care of the systematics. Great efforts are therefore ongoing to both determine which are the main sources of systematics and to investigate their impact on the estimation of the cosmological parameters.

As a preliminary but yet fundamental step, we have here presented a general formalism to make a step\,-\,by\,-\,step derivation of the multiplicative bias and of the additive systematics power spectra originating from shape measurement errors and uncertainties on the PSF reconstruction. Moreover, the proposed algorithm explicitly takes into account the properties of the target galaxies so that the redshift evolution of the systematics is consistently computed rather than imposed a priori. To the best of our knowledge, this is the first time that the systematics power spectra are inferred from the survey design (entering through the PSF wavelength dependence and filter transmission curve) and the galaxies photometric properties.

Two quite general assumptions are at the core of our procedure. First, we have assumed that both the reconstructed PSF and measured galaxy shape moments can be related to the actual ones by linear relations (\ref{eq: shapebias}) and (\ref{eq: psfbias}). Rather than getting lost in the details of the PSF reconstruction procedure and shape measurement algorithm, we have parameterized all these systematics in the parameters $(m_{\xi}, c_{\xi}, m_{\omega}, c_{\omega})$ and $(\mu_{\xi}, \gamma_{\xi}, \mu_{\omega}, \gamma_{\omega})$ which can be set a posteriori once they have been determined (based on, e.g., a comparison with simulations) for the given measurement process. For the application presented in this work, we have estimated the $(\mu_{\xi}, \gamma_{\xi}, \mu_{\omega}, \gamma_{\omega})$ parameters considering the impact of the PSF wavelength dependence, while $(m_{\xi}, c_{\xi}, m_{\omega}, c_{\omega})$ have been given as function of the $S/N$ and apparent magnitude of the target galaxy. Although the details of our choice could be questioned, we are nevertheless confident that our algorithm offers the unique possibility to estimate the amplitude of the systematics and its redshift evolution according to the features of the different players entering the game hence representing a significant step forward towards a fully realistic description of these fundamental aspect of the cosmic shear analysis.

A possible caveat is represented by the use of unweighted moments in the derivation of Eqs.(\ref{eq: defmulbias})\,-\,(\ref{eq: defaddbias}).  In fact, in order to increase the S/N, one rather usually measures weighted moments; thus, the lensing shape transforms we have started from are not valid anymore in this case. This is actually not a serious shortcomings for our derivation. Indeed, as far as the relation between weighted and unweighted moments can be roughly approximated as linear, one can still use Eqs.(\ref{eq: defmulbias})\,-\,(\ref{eq: defaddbias}) provided $(m_{\xi}, c_{\xi}, m_{\omega}, c_{\omega})$ are consistently redefined. We have checked that this is indeed the case using a Gaussian weight function and varying its size with respect to the galaxy half light radius. For a wide range of galaxy properties and relative size, we indeed find a linear relation between weighted and unweighted moments so that we are confident that Eqs.(\ref{eq: defmulbias})\,-\,(\ref{eq: defaddbias}) and hence our algorithm for the systematics amplitude derivation can be safely trusted upon.

While our method represents a step forward for what concerns the amplitude of the systematics, the shape of their scale profile still remains to be set by hand. In order to parameterize our ignorance of this issue, we have proposed a multigaussian expansion allowing us to consistently derive the additive power spectra provided the typical scale of the systematics and the coefficients of the expansion are set. Although still a large degree of arbitrariness is left, this approach nevertheless allows to mimic a wide set of shape profiles, preserving a link with some quantities which can be inferred from an analysis of the possible sources of systematics. In order to move forward, one must unavoidably rely on simulations looking for a correlation of the shape measurement bias parameters $(m_{\xi}, c_{\xi}, m_{\omega}, c_{\omega})$ with the properties of the galaxy population. Let us assume, for instance, that they correlate with the color of the galaxy and hence with the morphology (early\,-\,type galaxies being typically redder than late\,-\,type ones). Since clustering properties of ellipticals are different  from those of spiral ones, one could then expect that the power spectrum of this kind of systematics is somewhat related to the galaxy ones, thus gaining a first hint on its shape profile. Unfortunately, a similar analysis is unavailable at the moment so that we are forced to leave free the coefficient of our multigaussian expansion.

The reliable procedure presented here to compute the systematics power spectrum provides us the necessary input to investigate their impact on the determination of cosmological parameters. Two effects are actually possible. Although much effort will be dedicated in future surveys to reduce as much as possible the sources of systematics, it is nevertheless possible that some of them will go undetected, so that the observed shear power spectra differ from the underlying cosmological one. One can therefore wonder to which extent fitting the data assuming that no systematics are present can bias the cosmological parameters. This is the point we have addressed here performing an MCMC analysis on mock data  with and without added systematics. We find that the best fit models trace the data at low multipoles, but poorly fits the high $\ell$'s, thus leading to very large $\chi^2$ values. This result suggests that a possible signature of the systematics is represented by the large $\chi^2$ values due to the poor performance of the best fit models at high multipoles, i.e. over a range where statistical errors are quite small, although the same effect could also be  provided rather by the need of a different theoretical model to describe the data. One could, however, decide to still trust the inferred cosmological parameters assuming that the large $\chi^2$ is only a consequence of the residual systematics. Such a choice will however lead to strongly biased constraints with the recovered parameters shifting from the input one many times the $1 \sigma$ error. This is in agreement with previous results in the literature (in particular with the outcomes of the M13 analysis) which is a reassuring test considering we have relied on a different and ameliorated derivation of the systematics multiplicative bias and additive power spectra and a MCMC fit to mock data instead of a Fisher matrix analysis.

As a further improvement to the procedure presented in this paper, one could be less optimistic and admit that some undetected systematics are present in the data. The point is now how to model them and to which extent fitting for their corresponding parameters degrades the constraints on the cosmological ones. In a sense, this is a complementary question to the previous one. We are now no more concerned in the bias, but rather in the accuracy. To this end, one should first find an analytical description of the systematics multiplicative term and additive power spectra. The procedure we have presented allows to work out a large range of realistic ${\cal{M}}(z)$ and ${\cal{A}}_{ij}(\ell)$ functions which can drive the choice of the better way to model the systematics. As Fig.\,\ref{fig: psplots} shows, the previous proposed parameterization available in the literature are unable to correctly mimic the systematics power spectra so that their use should be avoided. On the contrary, a more reliable strategy would be to generate a large set of ${\cal{M}}(z)$
and ${\cal{A}}_{ij}(\ell)$ functions and fit them with an analytical approximation flexible enough to mimic them, but with not too many parameters in order to avoid a severe weakening of the accuracy on the cosmological parameters. This will be subject of a forthcoming publication.

As a concluding remark, we would like to draw the weak lensing community attention on the need for a realistic derivation of the systematics power spectra. Notwithstanding which method is used to investigate the bias on the cosmological parameters (should it be based on a Fisher matrix approach or a MCMC fitting of mock data), the main point is to fully take into account all the players entering the game. As we have shown here, this means to realistically describe not only the shape measurement bias and the PSF reconstruction errors, but also the distribution and the evolution with redshift of the galaxy properties. It is only a combined analysis of all these factors which can finally lead us to avoid the most dangerous pitfalls undermining the potential of cosmic shear to put an end to the quintessence vs modified gravity fight for the dark energy throne.

\section*{Acknowledgements}

We warmly thank J. Amiaux, M. Cropper and C. Saxton for making the Euclid PSF available and H. Hoekstra, T. Kitching, R. Massey and S. Pandolfi for useful comments on an earlier version of the manuscript. VFC also ackwnoledges G. Coppola, A. Graham, J. Liske and C. Tortora for help with the galaxy catalog simulation. VFC and ZH are grateful to M. Lombardi and M. Radovich for discussions on the wavelength dependence of the shape parameters and for checking the formulae in Appendix A. VFC is funded by the Italian Space Agency (ASI) through contract Euclid\,-\,IC (I/031/10/0). VFC and RS acknowledges financial contribution from the agreement ASI/INAF/I/023/12/0. MM acknowledges partial support from the PD51 INFN grant.

\appendix

\section{Shape transforms}

Our derivation of the multiplicative and additive bias in Eqs.(\ref{eq: defmulbias}) and (\ref{eq: defaddbias}) relies on the use of the $(\xi, \omega)$ shape parameters. These are slightly different from the conventional ones $(\varepsilon, R)$ so that we find it useful to briefly sketch the derivation of the effect of lensing, PSF convolution and $\lambda$ integration in the moments space and find out how the measurable values of $\xi$ and $\omega$ relate to the reduced shear.

\subsection{Lensing transform}

As well known, lensing provides an achromatic mapping between the image and the lens planes. To the first order, this mapping can be linearized as\,:

\begin{displaymath}
x_i^s = A_{ij}(\vcx) x_j
\end{displaymath}
where $\vcx$ ($\vcx^s$) are the coordinates in the image (source) plane and the Jacobian matrix ${\cal{A}}(\vcx)$ is related to the lensing deflection potential $\psi(\vcx)$ as $A_{ij}(\vcx) = \delta_{ij} - \psi_{,ij}(\vcx)$ and is usually parameterized in terms of the convergence $\kappa$ and the reduced shear $g = \gamma/(1 - \kappa)$. The inverse mapping will be

\begin{displaymath}
x_i = A^{-1}_{ij}(\vcx) x_j^s \ ,
\end{displaymath}
while the area elements transforms as $dx_1 dx_2 = {\rm det} {A^{-1}} dx_1^s dx_2^s$. Because of the achromatic nature of the lensing phenomenon, the Liouville theorem ensures that the 2D energy distribution is conserved so that $E(\vcx, \lambda) = E(\vcx^s, \lambda)$. Using this relation and the above geometric transformations, it is then possible to get the following rule for the moments\,:

\begin{equation}
Q_{ij}(\lambda) = A^{-1}_{im} Q^0_{mn}(\lambda) A^{-1}_{nj} \ ,
\label{eq: qijlens}
\end{equation}
so that it is only a matter of algebra to finally get

\begin{equation}
\xi(\lambda) = \frac{2 g \omega_0(\lambda) + \xi_0(\lambda) + g^2 \xi_{0}^{\star}(\lambda)}{(1 - \kappa^2)(1 - |g|^2)^2} + \xi_{PSF}(\lambda) \ ,
\label{eq: xiobs}
\end{equation}

\begin{equation}
\omega(\lambda) = \frac{(1 + |g|^2) \omega_0(\lambda) + 2 {\cal{R}}(g \xi_0^{\star}(\lambda))}{(1 - \kappa^2)(1 - |g|^2)^2} + \omega_{PSF}(\lambda) \ ,
\label{eq: omegaobs}
\end{equation}
which, for $\gamma, \kappa << 1$, indeed reduce to Eqs.(\ref{eq: xiwl})\,-\,(\ref{eq: omwl}).

\subsection{PSF convolution}

For a given wavelength $\lambda$, the energy distribution $E(\vcx, \lambda)$ of a galaxy after convolution with a $\lambda$\,-\,dependent PSF with energy distribution $E_{PSF}(\vcx, \lambda)$ reads\,:

\begin{displaymath}
E(\vcx, \lambda) = \int{E_0(\vcx, \lambda) E_{PSF}(\vcx - \vcx^{\prime}, \lambda) dx_1 dx_2} \ .
\end{displaymath}
In the moments space, the convolution becomes a linear operation so that the following relation holds\,:

\begin{equation}
Q_{ij}(\lambda) = Q_{ij}^0(\lambda) + Q_{ij}^{PSF}(\lambda)
\label{eq: qijpsf}
\end{equation}
where, without loss of generality, we have assumed that, at each given $\lambda$, the PSF is centred on the galaxy. It is then not surprising that the ellipticity and size of the galaxy after convolution will be simply given by\,:

\begin{equation}
\xi(\lambda) = \xi_0(\lambda) + \xi_{PSF}(\lambda) \ ,
\label{eq: xipsfapp}
\end{equation}

\begin{equation}
\omega(\lambda) = \omega_0(\lambda) + \omega_{PSF}(\lambda) \ .
\label{eq: omegapsfapp}
\end{equation}
Note that, because of the $\lambda$ dependence, the impact of the PSF will be different at different wavelengths.

\subsection{Wavelength integration}

In common applications, the galaxy is imaged through a broadband filter so that what one measure is the wavelength integrated energy distribution, i.e. the intensity $I(\vcx)$ defined above. It is then possible to correspondingly define the integrated moments as\,:

\begin{equation}
Q_{ij} = \frac{\int{I(\vcx) x_i x_j dx_1 dx_2}}{\int{I(\vcx) dx_1 dx_2}} \ \quad \ (i, j = 1, 2)
\label{eq: qijint}
\end{equation}
and the $\lambda$\,-\,integrated ellipticity and size as\,:

\begin{equation}
\xi = Q_{11} - Q_{22} + 2 {\rm i} Q_{12} \ ,
\label{eq: defxiint}
\end{equation}

\begin{equation}
\omega = Q_{11} + Q_{22} \ .
\label{eq: defomega}
\end{equation}
Comparing $Q_{ij}$ and $Q_{ij}(\lambda)$ immediately gives\,:

\begin{equation}
Q_{ij} = \int{Q_{ij}(\lambda) {\cal{S}}(\lambda) {\cal{T}}(\lambda) d\lambda}
\label{eq: qijvsqijlambda}
\end{equation}
with ${\cal{S}}(\lambda)$ the flux normalized SED, i.e. $\int{{\cal{S}}(\lambda) {\cal{T}}(\lambda) d\lambda} = 1$. It is then easy to get\,:

\begin{equation}
\xi = \int{\xi(\lambda) {\cal{S}}(\lambda) {\cal{T}}(\lambda) d\lambda} \ ,
\label{eq: xivsxilambda}
\end{equation}

\begin{equation}
\omega = \int{\omega(\lambda) {\cal{S}}(\lambda) {\cal{T}}(\lambda) d\lambda}
\label{eq: omegavsomegalambda}
\end{equation}
thanks to the linearity of the $\lambda$ integration. Similarly, one can now integrate over $\lambda$ Eqs.(\ref{eq: xipsf})\,-\,(\ref{eq: omegapsf}) to show that the ellipticity and size of a galaxy convolved with a wavelength dependent PSF and observed with a broadband filter are still given by Eqs.(\ref{eq: xipsf})\,-\,(\ref{eq: omegapsf}) provided $[\xi_0(\lambda), \omega_0(\lambda)]$ are replaced by their $\lambda$\,-\,integrated counterparts and the PSF quantities are defined as\,:

\begin{equation}
\xi_{PSF} = \int{\xi_{PSF}(\lambda) {\cal{S}}(\lambda) {\cal{T}}(\lambda) d\lambda} \ ,
\label{eq: xivsxilambdapsf}
\end{equation}

\begin{equation}
\omega{PSF} = \int{\omega_{PSF}(\lambda) {\cal{S}}(\lambda) {\cal{T}}(\lambda) d\lambda} \ .
\label{eq: omegavsomegalambdapsf}
\end{equation}
It is worth stressing that the effective PSF ellipticity and size $(\xi_{PSF}, \omega_{PSF})$ are defined in terms of the normalized SED of the galaxy. As a consequence, the impact of the PSF will be different for galaxies having, e.g., the same ellipticity and size, but different stellar populations (hence, a different SED). This is an important point to keep in mind since it shows that it is not possible to define a unique effective PSF for an image, but rather the specific features of each galaxy have to be taken into account.

\section{Modelling galaxies}

A basic role in the computation of the multiplicative and additive bias is played by the galaxy structural parameters and its SED. It is therefore of paramount importance that the simulated galaxy catalog used to infer the systematics power spectrum is based on a realistic modelling of these galaxy properties. The way we choose the SED of each galaxy and details on how the bulge and disk structural parameters have been set and on their wavelength and redshift dependence are described below.

\subsection{Galaxy model and SED assignment}

A galaxy is modelled as a two component system made out by a Sersic (1968) bulge and an exponential \citep{F70} disk, i.e. the intensity profiles in a given filter $f$ read\,:

\begin{eqnarray}
I_b(x_1, x_2) & = & \frac{L_b b_n^{2n}}{2 \pi n R_{eff}^2 {\rm e}^{b_n} \Gamma(2n)} \\
~ & \times & \exp{\left \{ -b_n \left [ \left ( \frac{X_{b1}^2 + X_{b2}^2/q_b^2}{R_{eff}^2} \right )^{\frac{1}{2n}} - 1 \right ] \right \}} \ , \nonumber
\label{eq: ib}
\end{eqnarray}

\begin{equation}
I_d(x_1, x_2) = \frac{L_d}{2 \pi R_{d}^2} \exp{\left [ \frac{(X_{d1}^2 + X_{d2}^2/q_d^2)^{1/2}}{R_d} \right ]} \ ,
\label{eq: id}
\end{equation}
with

\begin{equation}
\left \{
\begin{array}{l}
X_{i1} = x_1 \cos{\theta_i} + x_2 \sin{\theta_b} \\
~ \\
X_{i2} = -x_2 \sin{\theta_i} + x_2 \cos{\theta_b} \\
\end{array}
\right . \ .
\label{eq: defcoord}
\end{equation}
A galaxy is then assigned by setting the value of the geometric quantities $(q_b, \theta_b, q_d, \theta_d)$ and the structural parameters for the bulge $(n, R_{eff}, L_b)$ and the disk $(R_d, L_d)$. Moreover, such quantities can be both wavelength and redshift dependent. We therefore first simulate a sample of galaxy at redshift $z = 0$ setting the parameters as summarized below.

\begin{itemize}

\item[(i.)]{We randomly generate the bulge position angle $\theta_b$ and then set the disk one as $\theta_d = \theta_b + \Delta \theta$ with $-25^o \le \Delta \theta \le 25^o$. The two components are taken to be concentric, while their axes ratios $(q_b, q_d)$ are generated according to the distribution given by \cite{Cr01}.} \\

\item[(ii.)]{In order to set the SED of the galaxy, we rely on the color\,-\,magnitude diagrams of the sample of local galaxies collected in the New York University Value Added Galaxy Catalog (NYU\,-\,VAGC, \citealt{B05}) based on the SDSS survey \citep{Y00}. After having randomly chosen a galaxy from the sample, we set the $(u - g, g - r, g - i, g - z)$ colors of the simulated galaxy equal to the average value of a sample of 2000 NYU\,-\,VAGC galaxies having $i$ mag within 0.1 of the starting one. We then assign to the simulated galaxy a SED chosen from a large library of single burst stellar population models\footnote{To this end, we use the {\tt Galaxev} code by \cite{BC03}. Note that we do not consider more complicated models in order to not boost the number of unknown parameters to be set.} obtained by varying the age, the metallicity and the bulge\,-\,to\,-\,total luminosity ratio ($B/T$). In order to choose the SED, we minimize the difference between the theoretical and observed colors over a 5D space
($B/T$ plus bulge and disk age and metallicity).} \\

\item[(iii.)]{Having thus chosen the SED of the galaxy, we can scale the bulge luminosity from the $i$ to the $B$ band and use a set of empirically motivated scaling relations (summarized in Appendix B and detailed in \citealt{C12}) to set the bulge and disk structural parameters. The luminosity of the two components and the value of their parameters in other wavebands are then fixed according to some simple rules inferred from observations of real galaxies.} \\

\item[(iv.)]{While points $(i.)$\,-\,$(iii.)$ allow to build up a sample of $z = 0$ galaxies in good agreement with the observed luminosity, color and color\,-\,magnitude diagrams, we are interested to $z > 0$ systems. To this end, we first assign a redshift $z$ to a galaxies sampling from the distribution\,:

    \begin{displaymath}
    p(z) \propto \left ( \frac{z}{z_0} \right )^{a} \exp{\left [ - \left ( \frac{z}{z_0} \right )^{b} \right ]}
    \end{displaymath}
    with $(a, b, z_0) = (2.0, 1.5, 1.412 z_m)$ and $z_m = 0.9$ the median redshift expected for the Euclid survey. We then use the galaxy SED to estimate the bulge and disk luminosity in the  $RIZ$ filter, while structural parameters have been scaled according to empirical relations (see Appendix B).} \\

\item[(v.)]{Finally, we include a galaxy in the catalog if and only if its apparent magnitude in the $RIZ$ filter $m_{RIZ}$ passes the selection cut $m_{RIZ} \le 24.5$ and its wavelength integrated ellipticity and size parameters have no anomalous values.}

\end{itemize}
The galaxy catalog thus constructed has the correct distribution for luminosity, colors and color gradient and correctly traces the evolution of the galaxy properties with  redshift. We are therefore confident that the inferred intrinsic ellipticity and size distributions are quite realistic and can be reliably used as input to the estimate of the mean  and variance of the $(m, b)$ bias parameters.

\subsection{Bulge parameters at $z = 0$}

The surface brightness profile of both early\,-\,type galaxies and the bulges of late\,-\,type ones are known to be well fitted by the Sersic (1968) law \citep{CCD93,GC97,PS97} so that we adopt this model for the red component of our simulated galaxies. As can be read from Eq.(\ref{eq: ib}), the Sersic model parameters are the total bulge luminosity $L_b$, the slope $n$ of the profile and the effective radius $R_{eff}$, while the position angle $\theta_b$ and the axial ratio $q_b$ sets the shape of the bulge. To this end, we start by choosing the Johnson $B$ band as reference and assign the luminosity $L_b$ as explained in Sect.\,4.1 above. The other parameters are set according to the following recipe.

\begin{itemize}

\item{{\it Position angle.} Assuming there is no preferred orientation (as it is expected for a field galaxy population), we randomly extract $\theta_b$ from a uniform distribution over the full $(0, 2\pi)$ range.} \\

\item{{\it Axial ratio.} Rather than generating $q_b$, we first sample the following distribution \citep{Cr01}\,:

\begin{equation}
P(\varepsilon) = \varepsilon \left [ \cos{\left ( \frac{\pi \varepsilon}{2} \right )} \right ]^2 \exp{\left [ - \left ( \frac{2 \varepsilon}{B} \right )^C \right ]}
\label{eq: ellpdf}
\end{equation}
with $\varepsilon = (1 - q_b^2)/(1 + q_b^2)$ and $(B, C) = (0.05, 0.18)$. Note that we manually cut the distribution at $\varepsilon = 0.9$ to avoid unrealistically flattened bulges. The bulge axial ratio is then obtained by solving for $q_b$ from the generated $\varepsilon$ value.} \\

\item{{\it Sersic index.} Following \cite{C09}, we assume that the Sersic index is correlated with the bulge absolute magnitude\footnote{Hereafter, we adopt the following convention that the underscript $b$ ($d$) will denote quantities referring to the bulge (the disk), while an upperscript $(f)$ labels the filter adopted.} ${\cal{M}}_{b}^{(B)}$ as

\begin{equation}
\log{n} = -0.1219 {\cal{M}}_{b}^{(B)} - 1.6829
\label{eq: nvsmb}
\end{equation}
with

\begin{displaymath}
{\cal{M}}_b^{(B)} = -2.5 \log{L_b} + B_{\odot}
\end{displaymath}
and $B_{\odot} = 5.33$ the Sun absolute magnitude in the $B$ band. Since the ${\cal{M}}_b^{(B)}$\,-\,$n$ correlation is affected by a $\sim 30\%$ scatter, for a given ${\cal{M}}_{b}^{(B)}$, we generate the Sersic index from a Gaussian distribution centred on the value predicted by Eq.(\ref{eq: nvsmb}) and a variance set equal to $30\%$ of the mean value.} \\

\item{{\it Effective radius.} It is well known that the effective radius of a galaxy is correlated with its luminosity. In particular, according to \cite{S03}, the $\ln{R_{eff}}$ distribution is well approximated by a Gaussian profile

\begin{eqnarray}
P(\ln{R_{eff}}) & = & \frac{1}{\sqrt{2 \pi} \sigma_{\ln{R_{eff}}}({\cal{M}}^{(B)})} \nonumber \\
~ & \times & \exp{\left \{ - \frac{1}{2}
\left [ \frac{\ln{R_{eff}} - \langle \ln{R_{eff}} \rangle({\cal{M}}^{(B)})}{\sigma_{\ln{R_{eff}}}({\cal{M}}^{(B)})} \right ]^2 \right \}} \nonumber
\end{eqnarray}
with

\begin{displaymath}
\langle \ln{R_{eff}} \rangle = -0.4 a_{eff} {\cal{M}}^{(B)} + b_{eff} \ ,
\end{displaymath}

\begin{displaymath}
\sigma_{\ln{R_{eff}}} = \frac{\sigma_1 - \sigma_2}{1 + {\rm dex}({\cal{M}}^{(B)} - {\cal{M}}_0)} \ .
\end{displaymath}
and we have defined ${\rm dex}(x) = 10^x$. It is worth noting that \cite{S03} actually fitted the observed galaxy surface brightness with a single Sersic profile and then separated the sample in early and late\,-\,type systems according to some selection criteria. As such, ${\cal{M}}^{(B)}$ is actually the absolute magnitude of the full galaxy and not of the bulge component only. However, since bulges share most properties of early\,-\,type galaxies (ETGs), we will use the above distribution setting $(a_{eff}, b_{eff}) = (0.65, -5.06)$ and $(\sigma_1, \sigma_2, {\cal{M}}_0) = (0.35, 0.27, -20.91)$ as found for ETGs.} \\

\end{itemize}
Having thus set the model parameters in the $B$ band, we now have to address their wavelength dependence. To this end, we rely on the literature to infer some approximate but nevertheless reasonable prescriptions. First, we will assume that the two geometric parameters $(\theta_b, q_b)$ are the same in all the filters. The SPIDER collaboration (\citealt{LaB10}, hereafter LaB10) has collected a large sample of low redshift galaxies and fitted single Sersic profiles to the surface brightness data in the $grizYJHK$ photometric bands. Fig.\,12 in LaB10 shows that the median and variance of the distribution of the axial ratio is almost the same over the full photometric system so that we are confident that our assumption is statistically well motivated. In the same paper, LaB10 also shows that the Sersic index and effective radius distributions are roughly consistent with each other, that is to say, the median and variance are almost the same along the full wavelength range covered by the $grizYJHK$ filters.
Actually, such a consistency is mainly due to the large variance of the distributions, while a marked trend of the median values is indeed present. We have then decided to fit these median values as function of the effective wavelength of the filter thus finding\,:

\begin{equation}
n^{(f)} \simeq n^{(B)} + 5.75 \times 10^{-5} \left [ \lambda_{eff}^{(f)} - \lambda_{eff}^{(B)} \right ] \ ,
\label{eq: nvslambda}
\end{equation}

\begin{equation}
\log{R_{eff}^{(f)}} \simeq \log{R_{eff}^{(B)}} - 9.32 \times 10^{-6} \left [ \lambda_{eff}^{(f)} - \lambda_{eff}^{(B)} \right ] \ ,
\label{eq: reffvslambda}
\end{equation}
with $\lambda_{eff}^{(f)}$ in nm and a root mean square of the residuals of order $10\%$ for both fits. To take care of this scatter, we then generate $(n, \log{R_{eff}})$ in a given band $f$ by randomly sampling Gaussian distributions centred on the above values and variance equal to the rms of the residuals.

\subsection{Disk parameters at $z = 0$}

Having thus fixed the bulge component, we now turn our attention to the disk one modelled with the standard exponential profile in Eq.(\ref{eq: id}) assuming there is no offset between the bulge and disk centres. The disk position angle is no more generated randomly, but is rather set as $\theta_d = \theta_b + \Delta \theta$ with $\Delta \theta$ randomly sampled between $(-25, 25) \ {\rm deg}$. Such a limitation has been imposed to avoid generating systems with a large misalignment between the bulge and the disk since they are quite unrealistic or a signature of barred systems which we are not interested in. The disk axial ratio is instead generated according to the same functional distribution in Eq.(\ref{eq: ellpdf}), but setting $(B, C) = (0.19, 0.58)$ as found by \cite{Cr01} for late\,-\.type systems.

The SED assignment procedure gives an estimate of the bulge\,-\,to\,total luminosity ratio so that the disk luminosity $L_d$ is automatically set after having chosen the total luminosity and the SED. In order to be sure that the total galaxy profile is realistic, we then rely on the bulge\,-\,disk decomposition of $\sim 10000$ galaxies of the Millennium Galaxy Survey \citep{Lis03} made by \cite{A06}. We select galaxies well fitted by the sum of a Sersic component and an exponential profile and with measured redshift $z \le 0.1$ and use this subsample (comprising $\sim 50\%$ of the full sample) to investigate the dependence of the disk parameters on the bulge ones. In particular, we have looked for a correlation between the the disk to bulge scalelength ratio\footnote{We neglect here the small difference between the Johnson $B$ filter used to define the bulge quantities and the $B_{MGC}$ filter of the Millennium Galaxy Survey.} $R_d/R_{eff}$ and the bulge parameters themselves. We indeed find that the relation

\begin{equation}
\log{\left ( \frac{R_d}{R_{eff}} \right )} = -0.0653 {\cal{M}}_b -0.8640 \log{R_{eff}} - 0.7715
\label{eq: rdreffratio}
\end{equation}
provides a reasonable well fit to the data with a rms scatter of the residuals $\sim 30\%$. We therefore extract the quantity $\log{(R_d/R_{eff})}$ from a Gaussian distribution with centre and variance defined by our fit and then use this value to set $R_d$ given the bulge effective radius.

While the above scaling relations allow us to set the disk parameters in the $B$ band, we have still to decide how to assign them in other filters. To this end, we first scale the disk luminosity according to the SED stressing that, since the scaling is not the same for bulge and disk, the bulge luminosity fraction $f_b$ will be wavelength dependent too. This is consistent with the common sense intuition that $f_b$ will be larger in the redder wavebands being the bulge made out of an older population than the disk. In order to scale the disk scalelength radius $R_d$, we should have a model for the wavelength dependence of the $R_d/R_{eff}$ ratio. Unfortunately, we can not rely on the Millennium Galaxy Catalogue since this is a monochromatic survey so that we will make the rough assumption that the ratio $R_d/R_{eff}$ is constant within $25\%$ over the wavelength range covered by the SDSS and Euclid filters we have used to generate the catalog. As a consequence, we will set $R_d$ in the other bands randomly
generating $\log{(R_d/R_{eff})}$ from a Gaussian distribution centred on the $B$ band value and with a variance set to $25\%$  of this value. We then set $R_d^{(f)}$ multiplying the sampled $R_d/R_{eff}$ by the bulge effective radius in the filter of interest, evaluated using the scaling introduced above.

\subsection{Redshifting structural parameters}

The above procedure allows to set, in a realistic way the bulge and disk structural parameters for a galaxy at $z = 0$. Actually, the galaxies in the simulated catalog are not local ones so that we need a procedure to relate the present day parameters to the their high $z$ counterpart. While this is quite easy for the total luminosities using the assigned SED, redshifting the galaxy components back in time has also to take into account whether and how the structural parameters evolve. To this end, we adopt the strategy we briefly sketch below.

\begin{itemize}

\item{{\it Sersic index.} In a hierarchical formation scenario, ETGs may come out from the merging of two LTGs. From the point of view of galaxy modelling, we therefore expect that, as $z$ gets larger, we find galaxies described by a single Sersic component with a median index $n$ becoming closer to the disk value ($n = 1$) as we go back in time. Tracking the redshift dependence of $n$ is actually a problematic task since estimating $n$ for galaxies at high $z$ is quite hard given that one has to fit a three\,-\,parameters model to data covering only a very limited range. As such, one has first to check whether the fitting procedure is reliable or not and then can rely on the estimated $n$ to investigate the dependence on $z$. Notwithstanding all these caveats, some studies of limited samples can be found in literature \citep{Chi09,Cas10,Sz11}. We have therefore collected the values reported in these papers and fit a power\,-\,law relation, $n(z) \propto (1 + z)^{\nu}$, finding that $\nu \simeq -0.87$ fits
the data with a rms residual of $\sim 15\%$. Such a scatter could be narrowed down binning galaxies according to their specific star formation rate or stellar mass, both quantities which are unavailable for our simulated galaxies. Although we are well aware that the power\,-\,law fit only provides a crude approximation (since we have also not fully corrected for the different rest frame bands probed), this simple scaling allows us to mimic the flattening of $n$ with $z$ which is expected in a scenario where the ETGs fraction decreases with the redshift. We will therefore set the Sersic index in the $B$ band at redshift $z$ extracting its value from a Gaussian distribution centred on $n^{(B)}(z = 0) (1 + z)^{-\nu}$ and with a variance equal to $15\%$ of the mean value. We then use the same scaling with $\lambda$ adopted for $z = 0$ to get the Sersic index in the other filters.} \\

\item{{\it Scalelengths.} Observations tell us that the scalelengths also evolve with $z$. We follow here \cite{T06} who have compiled a large catalog of galaxies observed by the SDSS, GEMS and FIRES surveys to study the evolution of the galaxy size over the redshift range $(0, 3)$. Separating the galaxies in two groups according to the value of the Sersic index $n$, they find that the evolution of the effective radius for high $n$ systems (which can be identified with ETGs) is well fitted by\,:

\begin{equation}
R_{eff}(z) = R_{eff}(z = 0) (1 + z)^{\alpha}
\label{eq: reffvsz}
\end{equation}
with $\alpha = -1.01 \pm 0.08$, while it is\,:

\begin{equation}
R_d(z) = R_d(z = 0) E^{\alpha}(z)
\label{eq: rdvsz}
\end{equation}
with $\alpha = -0.83 \pm 0.05$ for low $n$ systems (approximating LTGs). We use Eqs.(\ref{eq: reffvsz}) and (\ref{eq: rdvsz}) to scale the $B$ band bulge effective radius and the disk scalelength to the redshift of the galaxy. In order to simulate the scatter around the best fit lines, we set $\alpha$ randomly sampling from a Gaussian distribution with mean and variance given by the measured values. The same scaling is applied to all the filters, i.e., we assume that Eqs.(\ref{eq: reffvsz}) and (\ref{eq: rdvsz}) are invariant under a filter change. Although there is no empirical evidence in favor or against this assumption, we prefer to start from this zero order approximation since (at our best knowledge) there are no studies investigating the wavelength dependence of the redshift scaling of $R_{eff}$ and $R_d$.} \\

\end{itemize}
Using these observationally motivated prescriptions, we can redshift our simulated galaxies and then compute their shape parameters thus finally getting all the ingredients needed to estimate the multiplicative and additive bias.

\end{document}